\documentclass[twocolumn, pra, aps, longbibliography, superscriptaddress]{revtex4-2}
\usepackage{amssymb, amsmath, amsfonts}
\usepackage{tikz}

\usepackage[]{inputenc}
\usepackage{color}
\usepackage{amstext}
\usepackage{enumitem}
\usepackage{graphicx, bm, palatino}
\usepackage[colorlinks=true, linkcolor=blue, urlcolor=blue, citecolor=blue, pdfusetitle]{hyperref}
\usepackage{cleveref}
\usepackage{microtype}
\usepackage{mathrsfs}
\usepackage[percent]{overpic}

\newcommand{\doubleletter}[2][1]{
  \scalebox{#1}{\rlap{#2}\hskip0.16pt #2}%
}

\usepackage[sc]{mathpazo}

\usepackage{cleveref}
\usepackage[dvipsnames]{xcolor}
\usepackage[caption=false]{subfig}

\usepackage{times}
\usepackage{bbm}


\begin{document}
\title{Revealing the non-classicality of a molecular nanomagnet}

\author{Alessandra Cammarata}
\affiliation{Universit\`a degli Studi di Palermo, Dipartimento di Fisica e Chimica - Emilio Segr\`e, via Archirafi 36, I-90123 Palermo, Italy}

\author{Steve Campbell}
\affiliation{School of Physics, University College Dublin, Belfield, Dublin 4, Ireland}
\affiliation{Centre for Quantum Engineering, Science, and Technology, University College Dublin, Dublin 4, Ireland}

\author{Mauro Paternostro}
\affiliation{Universit\`a degli Studi di Palermo, Dipartimento di Fisica e Chimica - Emilio Segr\`e, via Archirafi 36, I-90123 Palermo, Italy}
\affiliation{Centre for Quantum Materials and Technologies, School of Mathematics and Physics, Queen's University Belfast, BT7 1NN, United Kingdom}

\begin{abstract}
Molecular nanomagnets are compounds characterized by a high-spin magnetic core that is protected by organic ligands. They have recently gained attention as potential quantum information carriers in solid-state quantum computing platforms, simultaneously exhibiting classical macroscopic properties and quantum features in light of their complex nature and configuration. Addressing the condition when they manifest unquestionable quantum behavior is key to guarantee their effectiveness as resources for quantum information processing. 
We address the quantumness of molecular nanomagnets using a recently formulated criterion [cf. Krisnanda {\it et al.}, Phys. Rev. Lett. {\bf 119}, 120402 (2017)] demonstrating that these systems exhibit an intrinsic quantum nature, as evidenced by their ability to generate and enhance quantum correlations between two non-interacting probes. Our analysis, which is performed addressing various dynamical regimes, paves the way to the design of experimentally viable tests of non-classicality in  multipartite registers consisting of ensembles of molecular nanomagnets.
\end{abstract}

\maketitle

Molecular nanomagnets, also known as single-molecule magnets (SMMs), consist of organometallic molecules featuring a high-spin magnetic core protected by organic ligands, often carboxylates. They have recently attracted considerable interest from experimentalists and theorists alike~\cite{Chiesa_2024} because they exhibit macroscopic properties -- such as magnetization hysteresis -- typical of well‑known bulk magnets, while at the same time displaying microscopic features like quantum tunneling~\cite{wernsdorfer2004quantum}.  Their structure, which facilitates the coexistence of classical and quantum behaviors, makes them very appealing for the study of phenomena occurring at the mesoscopic scale, i.e. at the boundary between classical and quantum physics. Moreover, they offer applications in quantum computing, particularly in quantum error correction (QEC), where they serve as multilevel single objects~\cite{Chiesa2020}; in high‑density data storage, where they can significantly increase storage capacities~\cite{EmersonKing2025_SoftMagneticHysteresis,EmersonKing2025}; in spintronics, to improve device performance and reduce energy consumption~\cite{PhysRevB.99.245404}; and in quantum sensing as magnetic sensors for detecting fields at the nanoscale~\cite{LATINO2025170621}.

Characterizing the regime of operations of an SMM, in order to dig into their non-classical character, is thus relevant beyond fundamental motivations. In this work, we address this task by leveraging a \textit{non-classicality criterion}~\cite{PhysRevLett.119.120402} according to which a system in a genuine quantum state can generate entanglement between suitably arranged non-interacting probes, while a classical state could not. This approach has already proven fruitful to infer the non-classical nature of photosyntetic organisms~\cite{Krisnanda2018}, to study intra-cavity mechanical systems, and to propose a way to test the potential quantum nature of the gravitational interaction~\cite{Krisnanda2020}. Here, we further extend the domain of applicability of such a criterion by addressing the possibility to witness genuine non-classical features in the state of an SMM through its interaction with the field of a multi-mode cavity. {We show that the latter, which do not interact directly with each other but only through the SMM, become entangled. The emergence of such entanglement therefore provides strong evidence of the non-classical nature of molecular nanomagnets.}

The remainder of the paper is structured in the following way. Sec.~\ref{section: overview of a single-molecule magnet} provides a brief review of the physical structure, main features, modeling and applications of SMMs. In Sec.~\ref{section:working setup} we construct a working setup -- based on placing an SMM in a multi-mode cavity -- design so as to apply the chosen non-classicality criterion. In Sec.~\ref{section:results and discussion} the core results stemming from this approach are presented and discussed, together with the techniques employed. {We begin by modeling the SMM as a system consisting solely of a high-spin magnetic core. After establishing a basic understanding of this simplified description, we refine the model by including the organic ligands and the surrounding solution, which are treated as a nuclear spin bath. Finally, previously neglected nonlinear terms are reintroduced and their effects on the system dynamics are analyzed. To assess the robustness of our results, we also develop an alternative approach based on a phenomenological density-matrix description.}
Finally, in Sec.~\ref{conc} we draw our conclusions and provide a forward look for potential further directions of investigation. Several technical details related to our results are reported in a set of Appendices.

\section{Overview and modeling of a single-molecule magnet}\label{section: overview of a single-molecule magnet}

Molecular nanomagnets typically feature one or several metal centers with unpaired electrons, forming polynuclear metal complexes surrounded by bulky ligands.
Prominent examples of single-molecule magnets include the dodecanuclear mixed-valence manganese-oxo cluster, $\operatorname{Mn}_{12}$, and the octanuclear iron(III) oxo-hydroxo cluster, $[\operatorname{Fe}_8\operatorname{O}_2(\operatorname{OH})_{12}(\operatorname{tacn})_6]^{8+}$, abbreviated as $\operatorname{Fe}_8$. Both systems exhibit a ground state with total spin $S = 10$, corresponding to the minimum-energy configuration of the individual spins forming the cluster \footnote{The $\operatorname{Fe}_8$ cluster consists of eight Fe(III) ions, each with spin $s = 5/2$. The total spin $S$ can be regarded as the vector sum of the individual spins. However, due to the competing ferromagnetic and antiferromagnetic interactions within the molecule, the spins arrange themselves so that the cluster attains a total spin of $S = 10$, corresponding to the minimum-energy configuration.}. Moreover, both systems are endowed with an Ising-type magnetic anisotropy, which stabilizes the spin states with $m = \pm 10$~\cite{WERNSDORFER20081086}. {In what follows, we focus on $\operatorname{Fe}_8$ which is computationally simpler to model and captures the salient features of our analysis. In Sec.~\ref{Extension} we briefly report results for $\operatorname{Mn}_{12}$, which are broadly in agreement with the analysis of $\operatorname{Fe}_8$, and highlight some interesting qualitative differences.}

The magnetic configuration of an SMM is mainly determined by exchange couplings between the ions that form its magnetic core and by their interactions with the crystal field. In particular, transition-metal based SMMs are generally made by several magnetic ions coupled together by exchange or super-exchange interactions.
Interestingly, below their blocking temperature, these molecules exhibit magnetization hysteresis, a classical macroscopic property of a magnet. Nevertheless, they also manifest quantum tunneling of magnetization and quantum phase interference, which are microscale properties. The simultaneous presence of classical and quantum effects makes SMMs particularly appealing to investigate. Another notable feature of these systems is the modular nature of their structure, enabling systematic modification or scaling of the molecule to tune magnetic properties, such as the blocking temperature~\cite{Duan2022}.

Having briefly outlined the main features of SMMs, we now construct a suitable model Hamiltonian to describe them.
In general, the physical properties of exchange-coupled transition-metal clusters are investigated using spin Hamiltonians. To model a molecular spin cluster, such as an SMM, two different types of Hamiltonian can, in principle, be used: the ``many-spin" Hamiltonian and the ``giant-spin" Hamiltonian. In the former, couplings between individual spin centers are considered, while in the latter the magnetic core of an SMM is treated as a single collective spin. Although the giant-spin Hamiltonian model does not explicitly account for single-ion anisotropies or their interactions, and can thus be considered as a less microscopic model, it remains relevant and provides significant insight into the physics of SMMs. Several studies have analyzed the mapping between the many-spin and the giant-spin Hamiltonians, and it has been shown that the two models are highly compatible as long as the temperature is low~\cite{GhassemiTabrizi2018}.

Indeed, the basic assumption behind the giant-spin Hamiltonian is that the molecule's ground spin state is well separated in energy from excited states, such that its magnetic properties can be described using a single-spin ground state. In other words, at low temperatures {($k_BT\!<\! \mathcal{J}$, with $\mathcal{J}$ the exchange coupling between the individual spins)} the magnetic behavior can be described by an effective total spin $S$ that results from the vector sum of the spins of the individual magnetic ions. Based on these observations, from now on the magnetic core of an SMM is modeled as a magnetic center composed of $n$ unpaired electrons with an associated ground state given by $S=n/2$. The carboxylate ligands, and the solution in which the nanomagnet is immersed, are modeled as a bath of surrounding nuclear spins~\cite{PhysRevB.77.054428}. In our model, nuclear spins are assumed to be protons $(I= 1/2)$, with $I$ the nuclear spin, since hydrogen nuclei typically represent the main source of decoherence. The Hamiltonian describing an SMM consists of several terms
\begin{equation}
    \label{Eq:1}
	H = H_{EZ} + H_{NZ}+H_{ZFS}+ H_{B} + H_{HF}.
\end{equation}
The first two terms correspond to the electron and nuclear Zeeman interactions, respectively; $H_{ZFS}$ is the crystal field (or zero-field splitting) Hamiltonian, $H_{B}$ is the bath Hamiltonian and, finally, $H_{HF}$ accounts for the hyperfine coupling with the bath. The explicit form of the Hamiltonian is
\begin{equation}
\begin{aligned}
H &=   \ \omega_e S_z  - \omega_b \sum_n {I}_n^z - D {S}_z^2 + E \left({S}_x^2 - {S}_y^2\right)\\& + \beta \sum_{n \neq k} \left( {I}_n^x {I}_k^x + {I}_n^y {I}_k^y \right) + \Delta \sum_{n \neq k} {I}_n^z {I}_k^z\\&+\alpha \left( {S}_x \sum_n {I}_n^x + {S}_y \sum_n {I}_n^y \right) + \gamma {S}_z \sum_n {I}_n^z.
\end{aligned}
\label{Eq:2}
\end{equation}
 Throughout this analysis, we adopt units such that $\hbar =1$. Here, $n\text{,} \ k=1 \text{,} \ ...  \text{,}\  \text{N}$, where $\text{N}$ is the total number of spins in the bath. The spin operator ${S}$ describes the magnetic core of the SMM, while the $n^{\text{th}}$-nuclear spin of the surrounding bath is denoted by $I_n$. The first two terms account for the interaction of the high-spin magnetic core and the nuclear bath with an external magnetic field, whereas the terms proportional to  the axial anisotropy constant $D$ and the transverse anisotropy constant $E$ describe the magnetic anisotropy of the nanomagnet.   The parameters $\beta$ and $\Delta$ determine whether the nuclear spins are strongly or weakly interacting with each other. We assume $\beta \text{,} \ \Delta \approx 10^3~\mathrm{Hz}$. Finally, the last two terms model the interaction between the magnetic core and the nuclear spin bath. In a solution, one typically assumes $\alpha=\gamma=1420~\mathrm{MHz}$~\cite{GoldfarbStoll2018}.

\section{Single-Molecule Magnet in \\ a multi-mode cavity}\label{section:working setup}
Quantum states are inherently fragile and the system of interest is often not directly accessible with the tools at our disposal. To overcome this limitation, we can shift our attention from the system itself to the way it interacts with external probes over which we have more control.
\begin{figure}[b!]
	\centering
	\includegraphics[width=0.68\columnwidth]{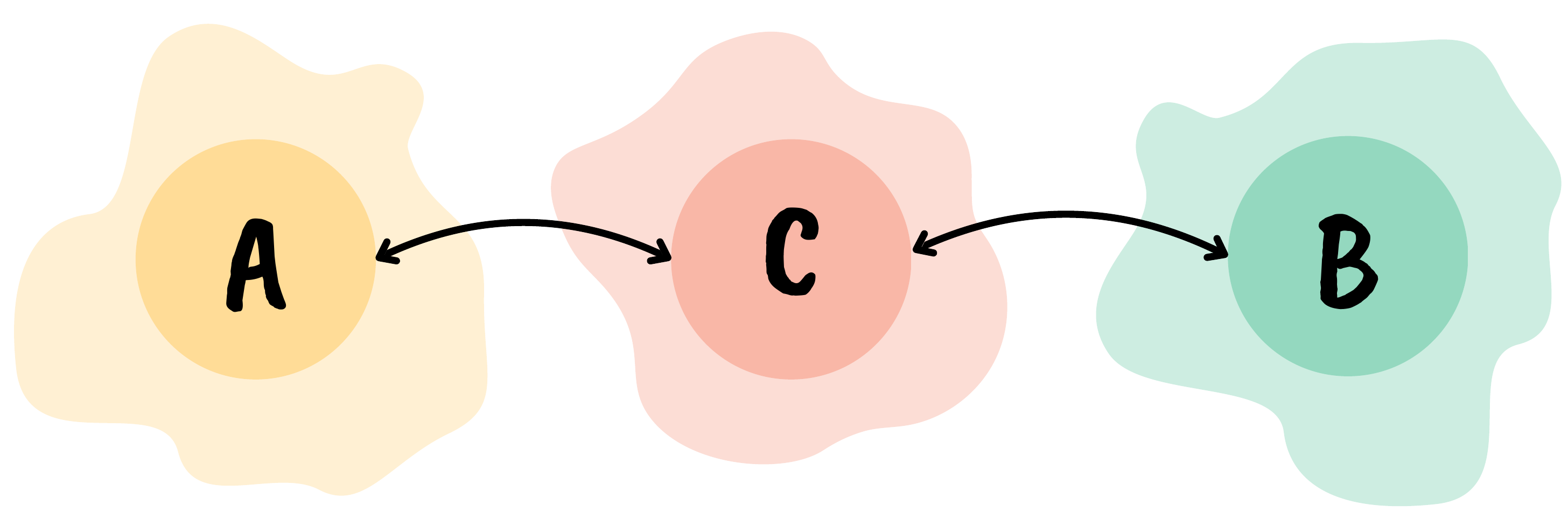}\\
    \vspace{5mm}
    \includegraphics[width=0.8\columnwidth]{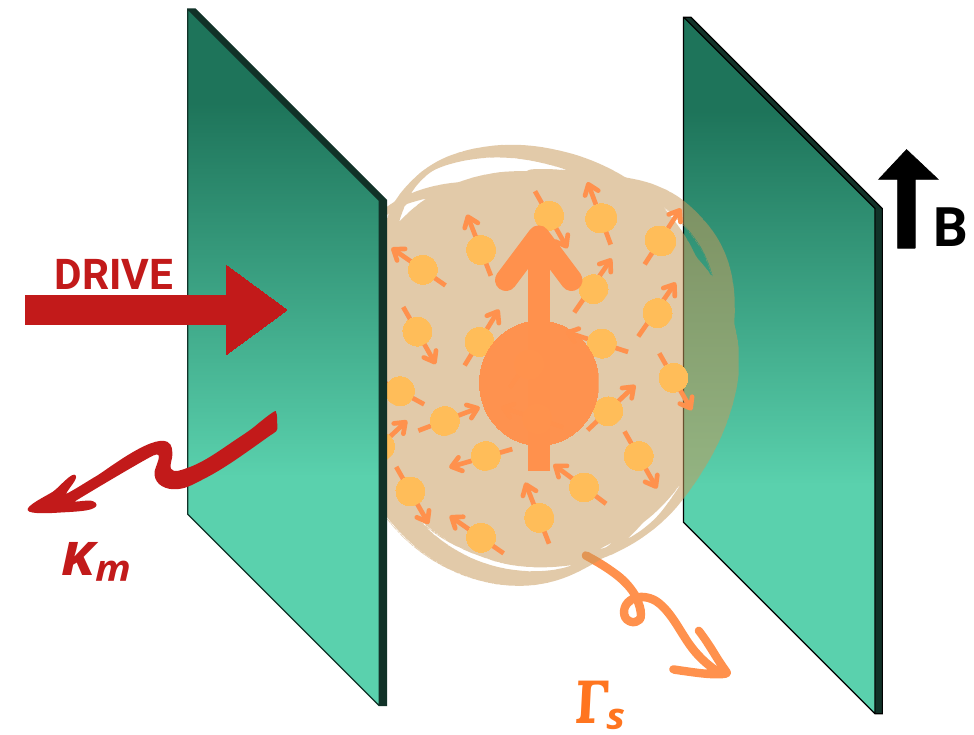}
    \captionsetup{font=small,labelfont={bf,small}}
	\caption{(Top panel)~Probes A and B individually interact with a mediator object C, but not with each other. Each system is open to its own environment, represented by the colored shadows. (Bottom panel)~Schematic representation of the working setup: an SMM is placed inside a multi-mode cavity. The straight black arrow indicates the presence of a magnetic field $B$, while the straight red arrow represents the external drive. The wavy red and orange arrows represent, respectively, the energy decay rates for the $m^{\text{th}}$-cavity mode and for the nanomagnet. Each cavity mode has a different frequency and they do not interact with each other.}
	\label{Fig:1}
\end{figure}
In this spirit, we introduce a scheme to infer the non-classicality of an inaccessible object based on its ability to mediate the creation of entanglement between controllable probe systems~\cite{PhysRevLett.119.120402}. Based on this criterion, it follows that the presence of entanglement implies that the inaccessible object must share non-classical correlations with the probe systems in form of quantum discord, thereby certifying its non-classical naturec. More concretely, consider three systems, labeled  A, B and C (see top panel of Fig.~\ref{Fig:1}). We will refer to C as the mediator; it represents our inaccessible object -- i.e., we neither have control over it nor can we perform measurements on it -- whereas A and B are the probes over which we have full control. 

An asset of this criterion is its generality. In fact, it can be applied to a wide range of different setups, requiring only a minimal set of conditions to work. Its versatility  emerges from the fact that the mediator may remain inaccessible at all times, i.e. it does not need to be directly measured. As a result, it can be either a very simple or a highly complex system. Furthermore, the details of the interaction between the mediator and the probes may also remain unspecified. Finally, every party can be open to its own local environment, so that the criterion is not restricted to closed systems.

To ensure the validity of the criterion, the following conditions must be met:
\begin{enumerate}[label=\arabic*.]
    \item The probes do not interact directly.
    \item The environment of each subsystem (probes and mediator) is local, so it does not interact with the other environments.
    \item The total initial state is a product state, so that no initial correlations exist between probes and mediator.
\end{enumerate}

Based on the above, we start from an initial product state and let the total system evolve. If, during the time evolution, the amount of entanglement between the probes becomes non-zero, then the criterion states that the mediator C cannot be described classically.

We now propose a feasible scheme for revealing non-classicality of an SMM and argue how it meets the three conditions above. Consider the arrangement shown in the bottom panel of Fig.~\ref{Fig:1}. The SMM, modeled as a giant-spin surrounded by a nuclear spin bath, is inside a driven multi-mode Fabry-Perot cavity where it interacts with some cavity modes. The latter are grouped into two sets that play the role of systems A and B in the general framework. The SMM embodies the mediator C.
The first requirement -- probes do not interact directly -- is easily met in a multi-mode cavity by considering modes with different frequencies and arbitrarily dividing them into two groups~\footnote{Alternatively, one could also exploit the polarization degree of freedom of light to generate two non-interacting sets of modes.}. 

The second requirement is more subtle. The electromagnetic environment of the modes lies outside the cavity, so it does not interact with the giant-spin or its environment. The only potential issue concerns the electromagnetic bath itself, which is common to all modes. We assume a large electromagnetic bath that induces the decay of each mode but without any back-action on them, i.e. each mode has a negligible impact on the bath. As a result, the information flowing from the modes to the environment is immediately lost. This ensures that different modes do not share information through the common bath and, consequently, do not interact via it. This is precisely the intended effect of the second assumption and therefore, under the conditions outlined above, the second requirement is satisfied.

Lastly, we require that, at the initial time, the total amount of entanglement is zero, i.e. the modes and the SMM are in a completely uncorrelated state at the initial time. This condition can be achieved in various ways. For example, one can prevent interaction between the SMM and the modes by applying a strong magnetic field. In fact, such a field modifies the energy level splitting of the SMM, detuning it from resonance with the modes.

The Hamiltonian describing the energy of such arrangement is

\begin{equation}
\begin{split}
    {H}&=\sum_{m} \omega_m {a}^\dagger_m {a}_m +\sum_{m} G ({a}^\dagger_m +{a}_m){S}_x+\omega_e {S}_z- D{S}_z^2\\&+ E({S}_x^2-{S}_y^2)- \omega_b \sum_{n}{I}_n^z + \gamma {S}_z\sum_n{I}_n^z+\Delta \sum_{n \neq k}{I}_n^z{I}_k^z\\&+\beta\sum_{n\neq k}\left({I}_n^x{I}_k^x+{I}_n^y{I}_k^y\right)+\alpha\frac{1}{2} \left( {S}_+\sum_{n}{I}_n^-+h.c.
    \right)\\&+ i\sum_m  E_m\left({a}_m^\dagger e^{-i\Lambda_m t}-h.c.
    \right).
    \label{Eq:3}
\end{split}
\end{equation}
Here, $m=1\text{,} ...\text{,}\ \text{M}$ is the label for the $m^{\text{th}}$-cavity mode (frequency $\omega_m$), whose annihilation (creation) operator is ${a}_m$ (${a}_m^\dagger$). $\text{M}$ is the total number of modes in the cavity. In this work, unless stated otherwise, we consider $\text{M}=6$. The second term represents the contribution arising from the giant-spin–photon interaction, whose strength is set by the coupling constant $G$. Our work is based on the interaction between the magnetic core of the SMM and the cavity modes, which  is far from trivial and deserves a more detailed discussion. Indeed, generating a strong spin-to-photon coupling is a challenging experimental task. This is because the interaction occurs via the magnetic field component, which results in values of $G$ that are several orders of magnitude weaker than those obtained from electric dipole-based couplings. However, recent progress allows us to consider $G =10 ~\mathrm{MHz}$, which we assume in our calculations~\cite{PhysRevApplied.19.064060, Samkharadze_2018}. 

The cavity is driven by a multi-mode laser with pumping frequencies $\Lambda_m=\omega_m$, and amplitude $E_m=\sqrt{2P_m\kappa_m/ \Lambda_m}$ (here, $P_m$ is the power for the $m^{\text{th}}$-pump and $\kappa_m$ the decay rate of the corresponding cavity mode). All remaining terms are associated with the nanomagnet and come from Eq.~\eqref{Eq:2}, in which we have expressed $S_x \text{,} \ S_y \text{,} \ S_z$ in terms of $S_+ \text{,} \ S_-$, and similarly for the nuclear spins.

We assume to have a Fabry-Perot cavity with fundamental frequency $\omega=6.75 \cdot 10^{11}~\mathrm{Hz}$, resonant with one of the allowed spin transitions of $\text{Fe}_8$.  The SMM is immersed in a diamagnetic solution with refractive index $n_r=\sqrt{\epsilon_r} \approx 1.33$. Thus, from $\omega_m=m\pi c/n_rL$, we deduce a cavity length of $L=1.05~\mathrm{mm}$. The reflectivities of the mirrors are taken as $R_1=100\%$ and $R_2=87\%$, and we assume them to be the same for all optical modes. Thus, using the expression for the finesse $\mathcal{F}=-2\pi/\ln(R_1R_2)=\pi c/2\kappa_m n_rL$, we obtain the decay rate of the modes as $\kappa_m \approx 7.5 \cdot 10^9~\mathrm{Hz}$. The assumption of equal decay rates for the different modes is realistic, as their frequencies are of the same order of magnitude~\footnote{For $\text{Mn}_{12}$: $\omega_m=1.64m\cdot10^{12}~\mathrm{Hz}$, $L=0.43\ \mathrm{mm}$, $R_1=100\%$, $R_2=94\%$ and $\kappa_m \approx 7.5 \cdot 10^9~\mathrm{Hz}$.}. In the case of the SMM, the decay rate is connected to relaxation and dephasing. However, at low temperatures, the relaxation time $(T_1)$ becomes several orders of magnitude longer than the dephasing one $(T_2)$. Therefore, we consider only the latter. Several investigations on spin dynamics estimate  $T_2$ in the range of nanoseconds~\cite{PhysRevLett.102.087603, Petiziol2021}, and we thus assume $\Gamma_s=1/T_2 \approx 10^{9}~\mathrm{Hz}$.


\section{Results and Discussion}
\label{section:results and discussion}
\subsection{Harmonic approximation}

To smoothly proceed with the analysis we introduce two key approximations: treating the nuclear spin bath as a single collective spin $J=\text{N}/2$~\cite{PhysRevA.81.032314} and assuming that the system operates at very low temperatures, allowing us to use the Holstein–Primakoff transformation~\cite{PhysRev.58.1098} (cf. Appendix~\ref{appendix1} for details).
Under such conditions, starting from Eq.~\eqref{Eq:3}, the resulting total Hamiltonian -- which we dub as the \textit{bosonic Hamiltonian} -- takes the form
{\small{
\begin{equation}
\label{Eq:4}
	\begin{aligned}
    &H =\sum_m\omega_m{a}_m^\dagger{a}_m+\Omega_s{s}^\dagger{s}+\Omega_n{n}^\dagger{n}- D({s}^\dagger{s})^2+\Delta({n}^\dagger{n})^2\\&+ ES({s}^2+h.c.)+G\sqrt{\frac S2}\sum_m({a}_m+{a}_m^\dagger)({s}+{s}^\dagger)-\gamma{s}^\dagger{s}{n}^\dagger{n}\\&+\alpha\sqrt{JS}({s}{n}^\dagger+h.c.) +i\sum_m E_m\left({a}_m^\dagger e^{-i\Lambda_mt}-h.c.\right)+\operatorname{const}.
    	\end{aligned}
\end{equation}}}
Here, ${s}$ (${s}^\dagger$) is the annihilation (creation) operator for the SMM giant-spin with frequency  $\Omega_s=2DS+\gamma J-\omega_e$, while ${n}$ (${n}^\dagger$) the annihilation (creation) operator for the nuclear spin bath with frequency $\Omega_n=\gamma S+2\beta J-2J\Delta-\omega_b$. We note that both $\Omega_s$ and $\Omega_n$ are positive, as the dominant contribution comes from the positive terms. Eq.~\eqref{Eq:4} contains quartic terms of the self-/cross-Kerr form~\cite{Shen_2022}, which need particular care. We thus proceed gradually by analyzing three different cases of increasing complexity:
\begin{enumerate}[label=\arabic*.]
    \item {\it Zeroth-order case}: Starting from Eq.~\eqref{Eq:4}, we neglect the bath-related and higher-than-quadratic terms. This approximation corresponds to considering only the magnetic core of the SMM, i.e. the electronic giant-spin.
    \item {\it First-order case}: We reintroduce the bath terms into the description and we focus on how the bath affects the evolution of the giant-spin. The inclusion of the nuclear spin bath enables the modeling of the carbon ligands and the embedding solution.
    \item {\it Second-order case}: The Kerr-like terms are included in the analysis and the total Hamiltonian is treated through a linearization process. This step allows for a more accurate characterization of the anisotropic magnetic nature of the SMM.
\end{enumerate}

In what follows, we will address each case quantitatively. 
\noindent

\begin{figure*}[t!]  
  \centering
 \makebox[0.5\textwidth][l]{\hspace{11mm}\doubleletter{A}}%
  \makebox[0.5\textwidth][l]{\hspace{11mm}\doubleletter{B}}\vspace{0.5mm}\\
  \includegraphics[width=0.51\textwidth]{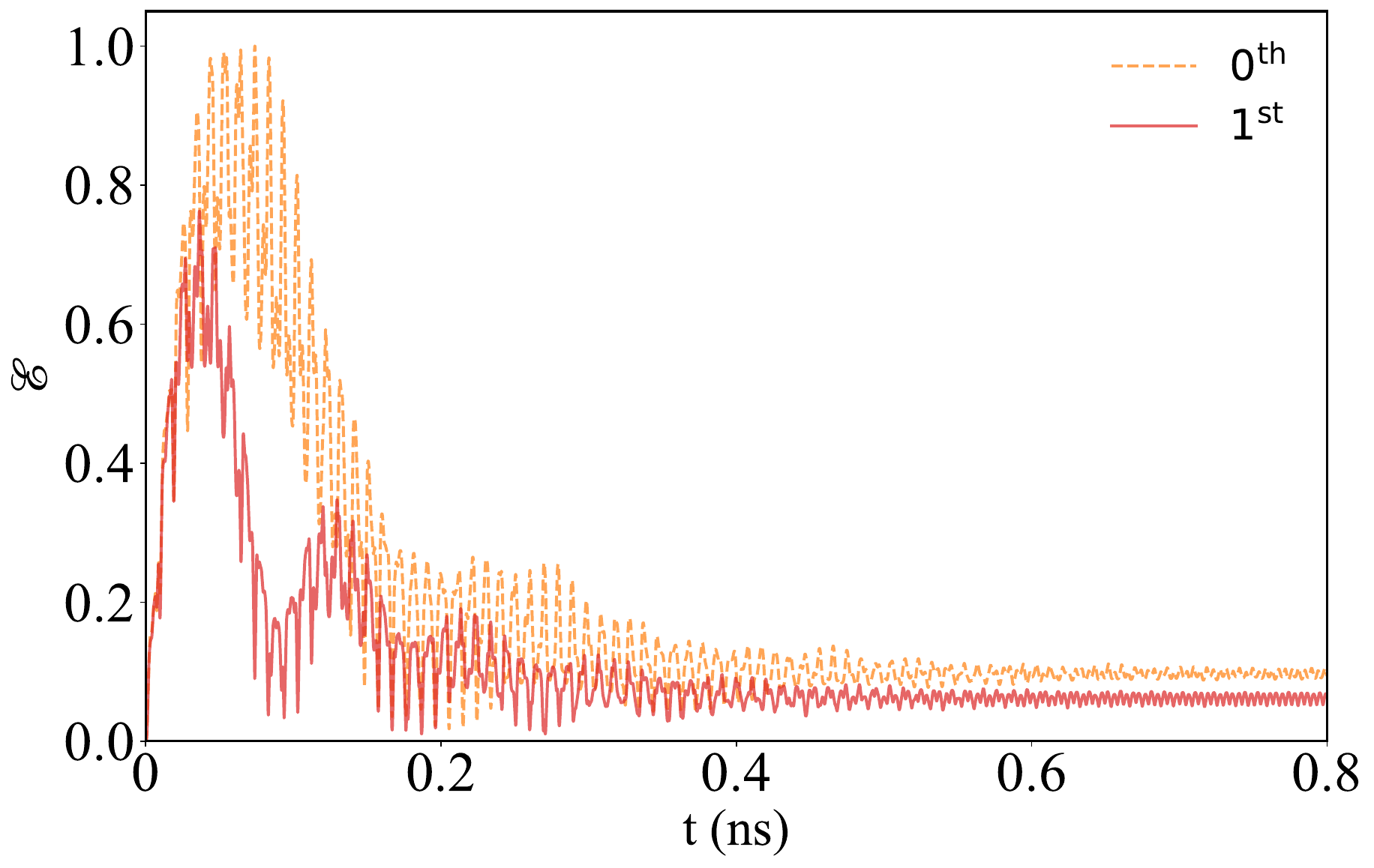}%
  \includegraphics[width=0.51\textwidth]{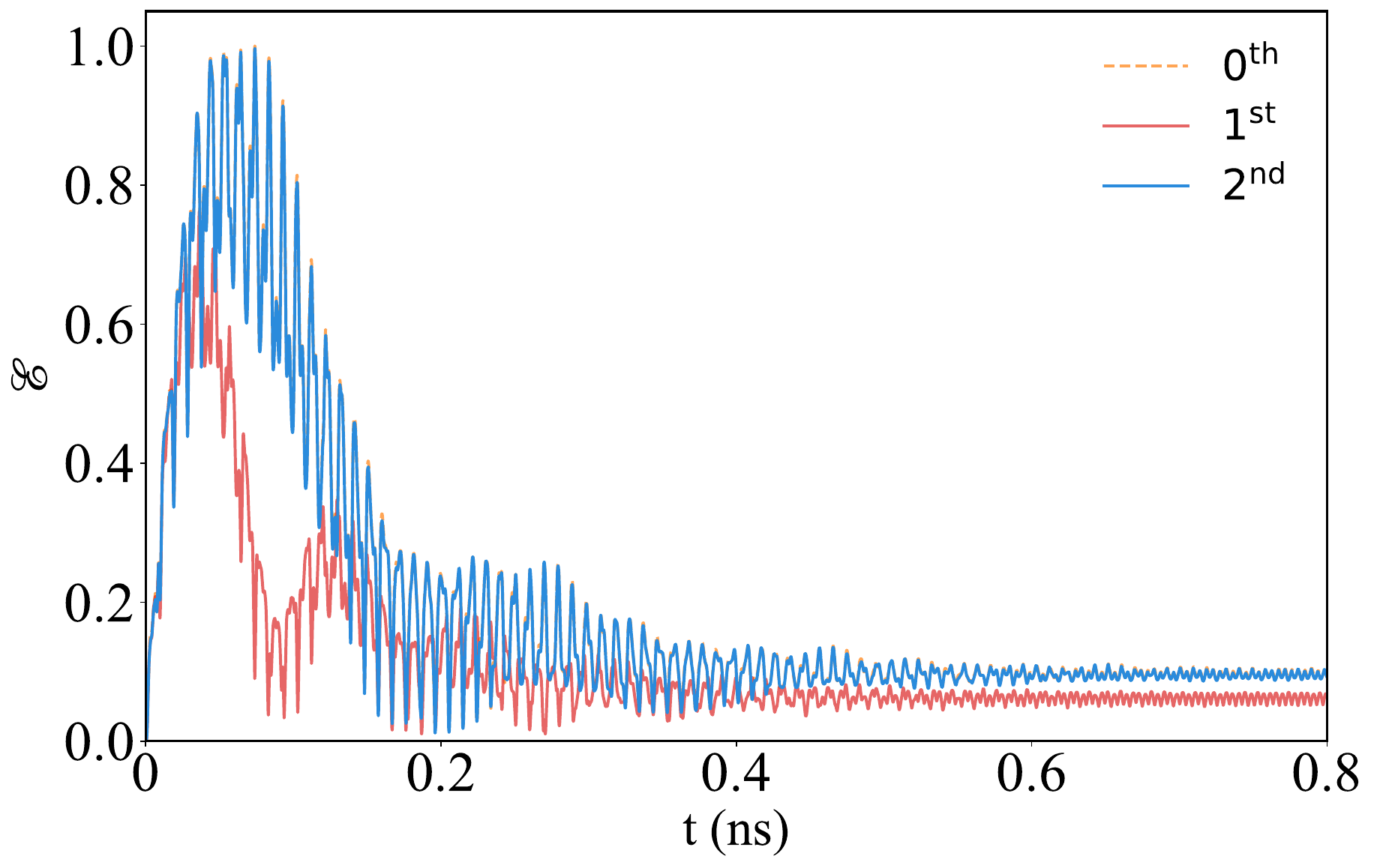}
    \vspace{0.3em}
 \makebox[0.5\textwidth][l]{\hspace{11mm}\doubleletter{C}}%
  \makebox[0.5\textwidth][l]{\hspace{11mm}\doubleletter{D}}
  \includegraphics[width=0.5\textwidth]{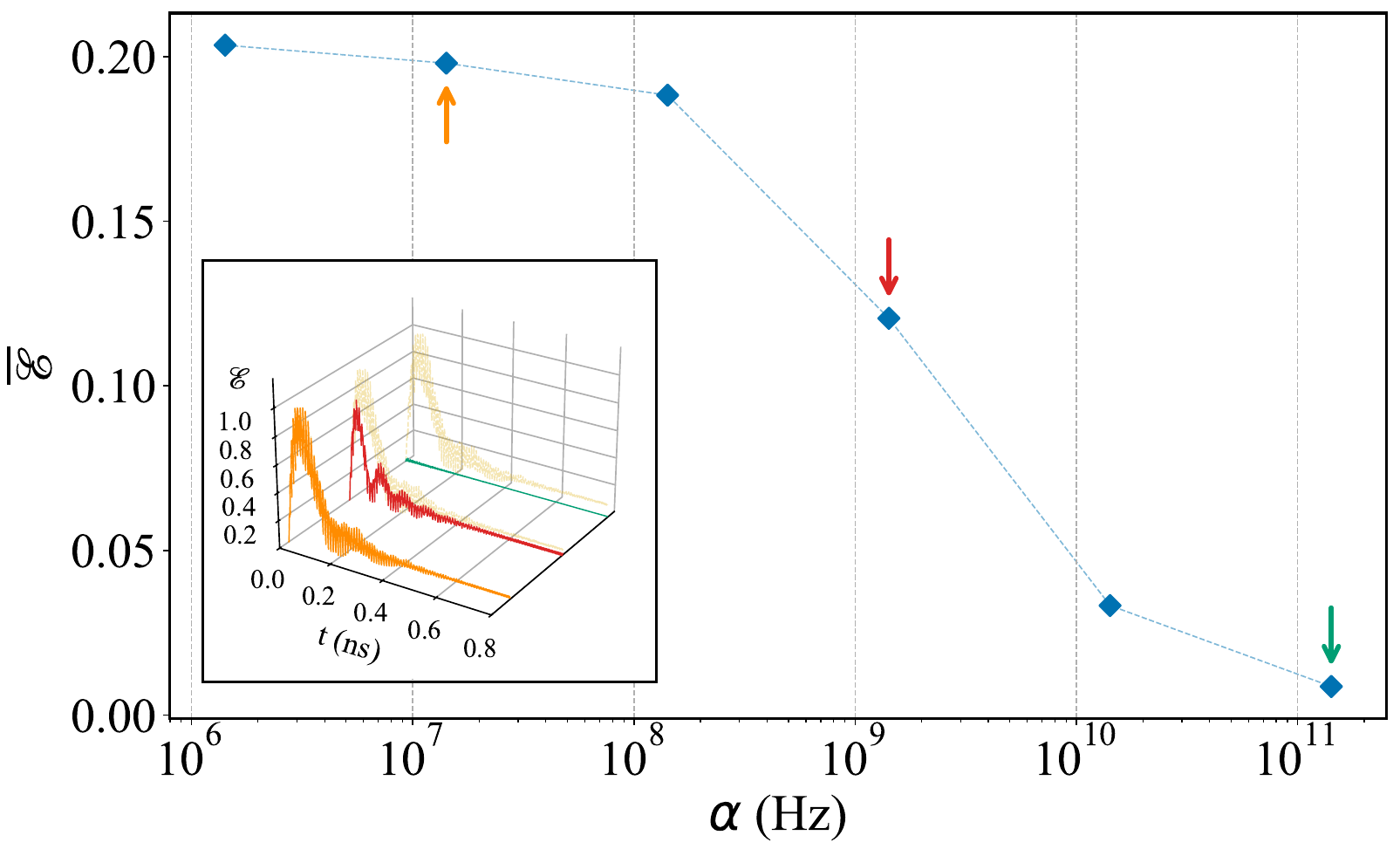}%
  \includegraphics[width=0.5\textwidth]{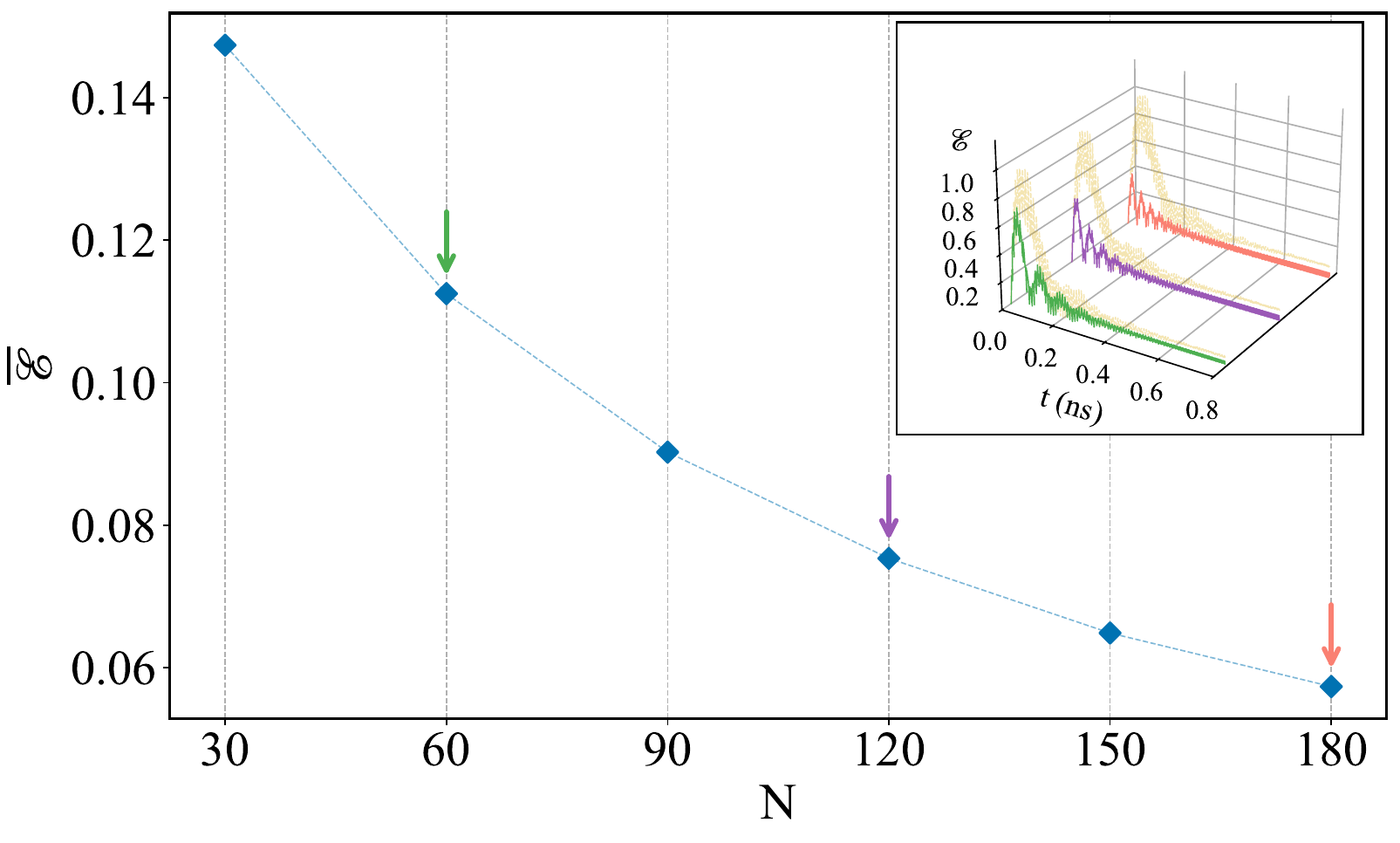}

  \captionsetup{font=small, labelfont={bf,small}}
  \caption{\!\!\doubleletter{A.} Bipartite entanglement in the zeroth-order case, without the effect of the bath, and in the first-order case, including the bath. The introduction of the environment leads to a reduction in entanglement without significantly altering its shape, as expected; \doubleletter{B.} Bipartite entanglement in the zeroth-order case, without the effect of the bath; in the first-order case, including the bath and in the second-order case, adding the non-linear terms with $K_1 =  3.6 \cdot 10^{10}~\mathrm{Hz}$ and $ K_2 = 1420 \cdot 10^6~\mathrm{Hz} $; \doubleletter{C.} First-order bipartite entanglement (time-averaged over the interval [$0$, $0.8]~\mathrm{ns}$) varying the hyperfine coupling strength: the environmental effect remains negligible until a certain value of $\alpha$ is reached; once this threshold is exceeded, the entanglement is rapidly destroyed. The 3D inset explicitly illustrates three representative cases of entanglement dynamics, the  yellow curve corresponds to the zeroth-order case, which remains unaffected by variation in the hyperfine coupling strength; \doubleletter{D.} First-order bipartite entanglement (time-averaged over the interval [$0$, $0.8]~\mathrm{ns}$) varying the number of spins in the bath. A higher number of spins in the bath increases the impact of the environment on the SMM, as expected. For all figures, without explicitly saying different, the parameters used are: $B = 0.01~\mathrm{T}\text{,}\ E = 6.02\cdot 10^9~\mathrm{Hz}\text{,}\ D = 3.6 \cdot 10^{10}~\mathrm{Hz}\text{,}\ S = 10 \text{,} \  \omega=6.75\cdot10^{11}~\mathrm{Hz} \text{,} \ J= 25  \ \text{and} \ \alpha=\gamma=1420 \cdot 10^6~\mathrm{Hz}.$ }
  \label{Fig:2}
\end{figure*}

\subsubsection{Zeroth-order case.}

While this regime might appear as based on too drastic a set of approximations, it is justified by the fact that the neglected terms are orders of magnitude smaller than the  frequencies of the harmonic oscillators for the giant-spin and for the modes. This clarifies why we refer to this approach as \textit{zeroth-order} case. The Hamiltonian takes the following form
 {\small{
 \begin{equation}
	\begin{aligned}
   &H_0 =\sum_m\omega_m{a}_m^\dagger{a}_m+\Omega_s{s}^\dagger{s}+ ES({s}^2+h.c.)\\&+\sum_m G\sqrt{\frac S2}({a}_m+{a}_m^\dagger)({s}+{s}^\dagger)+i\sum_m E_m({a}_m^\dagger e^{-i\Lambda_mt}-h.c.)\text{,}
	\end{aligned}
    \label{Eq:5}
\end{equation}}}
where $\Omega_s=2DS-\omega_e$. Eq.~\eqref{Eq:5} is Gaussian-preserving and allows for the swift derivation of the dynamical equations of motion, which can be cast in the form of the set of Langevin equations
\begin{equation}
	\dot{u}(t) = K u(t) + l(t).
	\label{Eq:6}
\end{equation}
Here  $u(t) = ({x}_1 \text{,} \ {y}_1 \text{,} \  {x}_2 \text{,} \ {y}_2 \text{,} \dots \text{,} \ {x}_\text{M} \text{,} \ {y}_\text{M} \text{,} \ {x}_s \text{,} \ {y}_s)^\mathrm{T}$ is the vector of $2(\text{M}+1)$ quadratures ${x}_m \equiv ({a}_m+{a}^\dagger_m)/ \sqrt{2} \text{,} \ {y}_m  \equiv ({a}_m-{a}^\dagger_m)/ i\sqrt{2} \text{,} \ {x}_s \equiv ({s}+{s}^\dagger)/ \sqrt{2} \text{,}\ {y}_s  \equiv ({s}-{s}^\dagger)/ i\sqrt{2}$ of the $\text{M}$ cavity modes and of the giant-spin, $K$ is the drift matrix, and $l(t)$  a vector containing both noise and driving terms. Matrix $K$ includes damping of the excitations of the giant-spin and cavity modes modes at rate 
$\Gamma_s$ and $\kappa_m$, respectively. In the following, unless stated otherwise, we set $\Gamma_s=0$. The effects of dephasing are explicitly accounted for by introducing the nuclear spin bath. The noise vector encodes the effects of zero-mean, delta-correlated Gaussian noise affecting the giant-spin and the cavity modes, here described by the operators ${F}$ and ${Q}_m$, respectively, with
\begin{equation}
    \langle {F}(t){F}(t')^\dagger\rangle=\delta(t-t')
\end{equation}
and 
\begin{equation}
    \langle {Q}_m(t){Q}_{m'}(t')\rangle =\delta_{mm'}\delta(t-t').
\end{equation}

\noindent 
The solution to  Eq.~\eqref{Eq:6} can be written as 
\begin{equation}
	u(t)=W_+(t)u(0)+W_+(t)\int_0^t dt'W_-(t')l(t')\text{,}
	\label{Eq:9}
\end{equation} 
where $W_{\pm}(t)=e^{\pm Kt}$. The system is stable and reaches a steady state when all the eigenvalues of $K$ have negative real parts such that $W(+\infty)=0$.  

From Eq.~\eqref{Eq:9}, we can derive an expression for the covariance matrix,~$V$ {whose entries are defined as 
\begin{equation}
	V_{ij}(t)=\frac{\langle \{\Delta u_i, \Delta u_j\}\rangle}{2}\text{,}
\end{equation}
where $\Delta u_i:= u_i-\langle u_i\rangle$ and $\{\cdot \text{,}\cdot\}$ is the anticommutator. The amount of entanglement among modes can then be quantified using the logarithmic negativity~\cite{Horodecki_2009,Plenio2005}
	\begin{equation}
	 	\operatorname{\mathscr{E}=\max}[0\text{,}-\ln(2\tilde\nu_{min})]\text{,}
        \label{Eq:10.1}
	\end{equation}
where $\tilde\nu_{min}$ is the smallest sympletic eigenvalue of the partially transposed covariance matrix $\tilde V$ encompassing the degrees of freedom of the cavity modes [cf. Appendices~\ref{appendix2}, \ref{appendix3} and \ref{appendix4} for details].}
{The analysis based on the zeroth-order Hamiltonian, shown in Fig.~\ref{Fig:2}\!\!\textcolor{blue}{\doubleletter{A}}, reveals a non-zero bipartite entanglement between the two groups into which we can partition the cavity modes.
Specifically, the $\text{M}\!=\!6$ modes are divided into two equally sized groups, with the partition chosen to maximize the inter-group entanglement. 

{Fig.~\ref{Fig:2}\!\!\textcolor{blue}{\doubleletter{A}} therefore establishes that the SMM exhibits non-classicality due to the creation of entanglement between the cavity modes. We remark that although the numerical value is small~\footnote{For the considered parameters, we find a maximum value of entanglement $\operatorname{\mathscr{E}}\!\propto\!10^{-8}$. Nevertheless, it is important to remark that this does not impact the physical interpretation, since the model is not designed to exploit quantum correlations but rather to assess whether the SMM generates any non-zero amount of entanglement between the probes.}, the key point is not the amount of entanglement, but only its presence. Therefore here, and in what follows, we renormalize the entanglement with the maximum value obtained dynamically simply to aid in the visualization.}
While this analysis suggests the SMM exhibits a genuinely non-classical nature, the model under consideration is not yet an accurate description of the situation at hand as it lacks several important terms. Therefore, we now proceed to refine such description by introducing the carbon ligands and the surrounding environment.}

\subsubsection{First-order case.} 
 We now introduce the bath-related terms in the Hamiltonian model of the nanomagnet-modes compound. We refer to this as the \textit{first-order} Hamiltonian, which takes the form
 \begin{equation}
H_1=H_0+\Omega_n{n}^\dagger{n}+\alpha\sqrt{JS}({s}{n}^\dagger+h.c.)
	\label{Eq:12}
\end{equation}
with $\Omega_s\to2DS+\gamma J-\omega_e$ [cf.~Eq.~\eqref{Eq:5}] and $\Omega_n=\gamma S+2\beta J-2J\Delta-\omega_b$. This Hamiltonian includes explicitly a term describing the energy of the spin bath and an interaction term between the nuclear-spin and the giant-spin, whose strength is set by the parameter $\alpha$. We point out that the addition of the environment causes a variation in the energy level spacing of the SMM's spin with respect to the zeroth-order case akin to the {Lamb shift}.

The Langevin equations stemming from Eq.~\eqref{Eq:12} can be derived and solved following lines similar to those presented previously [cf. Appendix~\ref{appendix5} for the corresponding detailed calculations]. 

As shown in Fig.~\ref{Fig:2}\!\!\textcolor{blue}{\doubleletter{A}}, the bath has a detrimental effect on the generation of entanglement. Nevertheless, the SMM is still able to produce a non-zero amount of entanglement between the modes. This indicates that, even in the presence of adverse environmental effects, the SMM maintains its non-classical character.

In Fig.~\ref{Fig:2}\!\!\textcolor{blue}{\doubleletter{C}} we analyze the effect of the bath on the entanglement by varying the coupling strength, $\alpha$. As observed in the simulations, the stronger the interaction with the bath, the lower the value of the observed entanglement. Indeed, the bath acts as a source of decoherence for the electronic spin. However, we highlight that the presence of the bath does not significantly affect the entanglement until the coupling strength exceeds a certain threshold value. This threshold appears to be related to the order of magnitude of the two anisotropy constants -- $D$ and $E$ -- which is reasonable since the entanglement is closely tied to the anisotropic nature of the SMM. However, it would be worthwhile to extend the analysis to a range of SMMs to determine whether this behavior is generally observed. This demonstrates a robustness of the entanglement against the decoherence effects introduced by the bath. 

In typical experiments, the number of nuclear spins in the bath can range from a few tens to several hundreds. For our simulations, we work with a bath consisting of 50 spins. As shown in Fig.~\ref{Fig:2}\!\!\textcolor{blue}{\doubleletter{D}}, increasing the number of nuclear spins leads to a gradual reduction in entanglement, in contrast to the sharper decrease observed when varying the coupling strength.


\subsubsection{Second-order case.}
We now compute the bipartite entanglement including environmental and non-linear effects in our analysis. We refer to this stage as \textit{second-order} case since, with respect to the first-order case, there is the addition of non-linear terms. Therefore, the Hamiltonian under study is
\begin{equation}
    H_2= H_1 - K_1({s}^\dagger{s})^2- K_2{s}^\dagger{s}{n}^\dagger{n},
    \label{Eq:13}
\end{equation} 
which gives rise to non-linear equations of motion due to the higher-order terms. Note that, according to Eq.~\eqref{Eq:4}, $K_1 \equiv D$ and $K_2 \equiv \gamma$, relabeled here to emphasize their role. 

The Hamiltonian in Eq.~\eqref{Eq:13} is quartic, which implies Gaussianity of the initial state of the system will no longer be conserved. However in order to gather insight without the need for the full-fledged simulation of the corresponding non-linear dynamics, we assume a modest contribution from the quartic terms and linearize the corresponding Langevin equations by decomposing each operator into a mean value and a (small) fluctuation term as  ${s}\approx  \langle s(t) \rangle+\delta {s}$, ${a}\approx  \langle a(t) \rangle+\delta {a}$ and ${n} \approx \langle n(t)\rangle+\delta{n}$. This is justified by requesting that the cavity is strongly driven, so as to macroscopically populate it and, in turn, achieve a strong polarization of the giant spin.
Using such linearized operators in the Langevin equations, which can be solved self-consistently by first determining the mean amplitudes of the modes $\langle s \rangle \text{,} \ \langle n \rangle \text{,} $ and $ \langle a_m \rangle$, the dynamics of the covariance matrix of the fluctuations can be predicted and, with that, the degree of entanglement that they share. Further details are provided in Appendix~\ref{appendix6}.

The three different cases analyzed so far are compared in Fig.~\ref{Fig:2}\!\!\textcolor{blue}{\doubleletter{B}}. The inclusion of the non-linear terms appears to counterbalance the detrimental effect of the bath, leading to a result that matches the zeroth-order case. One could argue that the structure of the SMM itself acts as a form of protection. However, this is not a generic feature of SMMs as it does not occur in the case of $\text{Mn}_{12}$ (see Sec.~\ref{Extension}), where the zeroth-order and second-order results differ. More generally, although the effects of non-linearity are difficult to predict, the results show that the inclusion of non-linear terms  has a positive incremental effect. This is an important finding, as the second-order case more accurately describes the nanomagnet. Therefore, the observation of entanglement, even in this case, provides further confirmation of the inherently non-classical nature of SMMs.
\begin{figure*}[t!]  
    \centering
    \makebox[0.33\textwidth][l]{\hspace{8mm}\doubleletter[0.92]{A}}%
    \makebox[0.333\textwidth][l]{\hspace{6mm}\doubleletter[0.92]{B}}%
    \makebox[0.333\textwidth][l]{\hspace{6.5mm}\doubleletter[0.92]{C}}\vspace{0.4mm}\\
    \includegraphics[width=0.68\columnwidth]{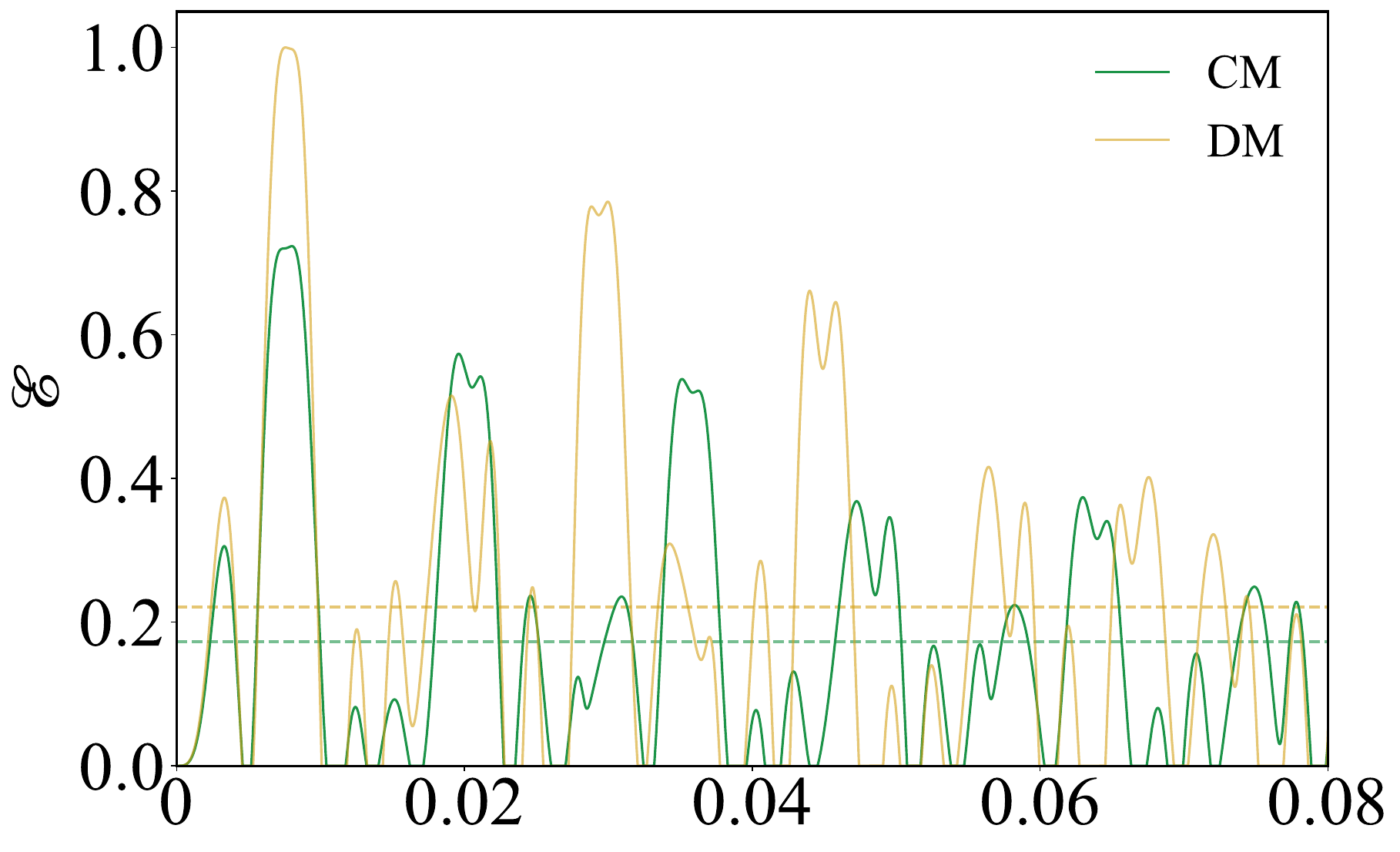}
    \hfill
    \includegraphics[width=0.68\columnwidth]{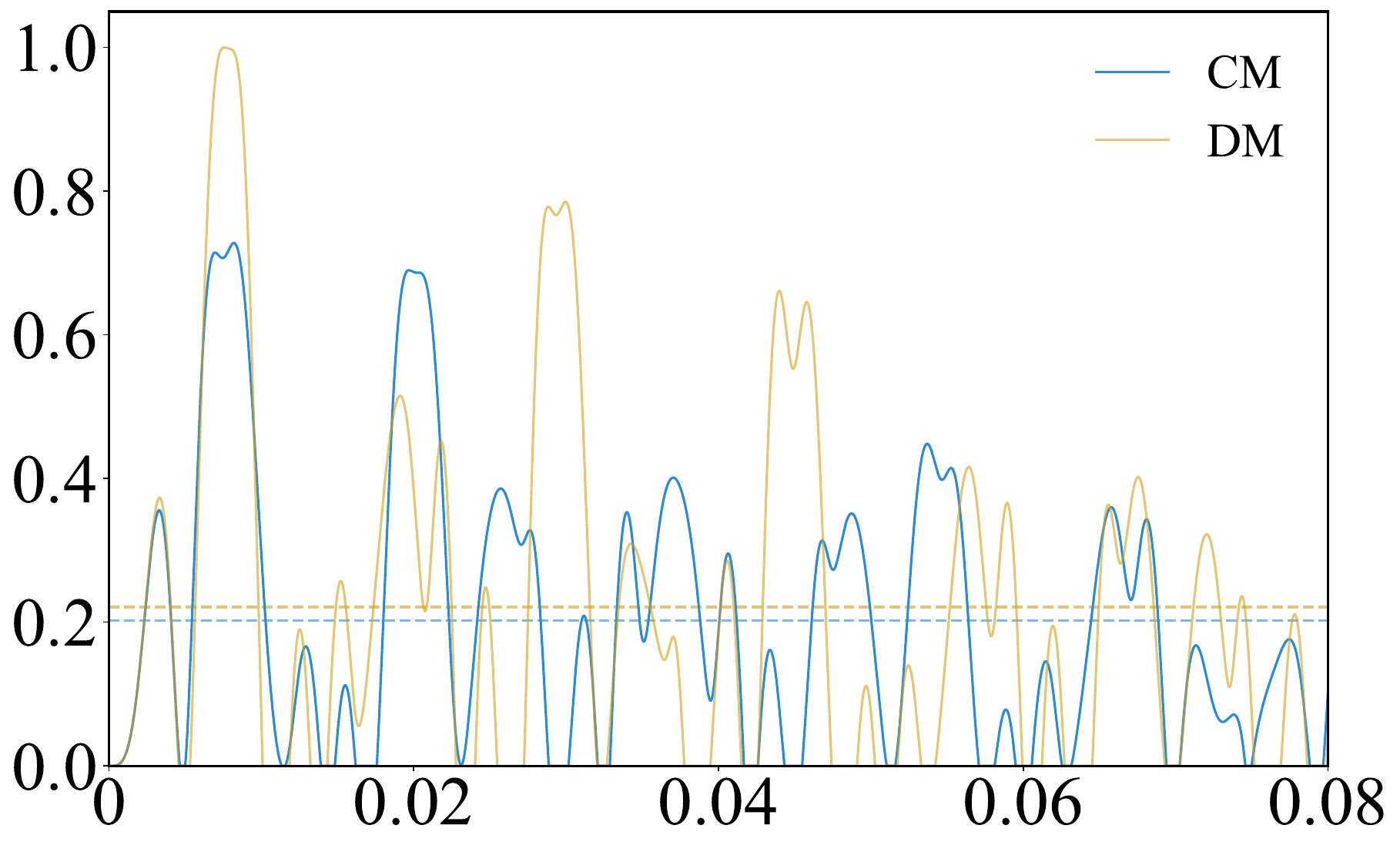}
    \hfill
    \includegraphics[width=0.68\columnwidth]
    {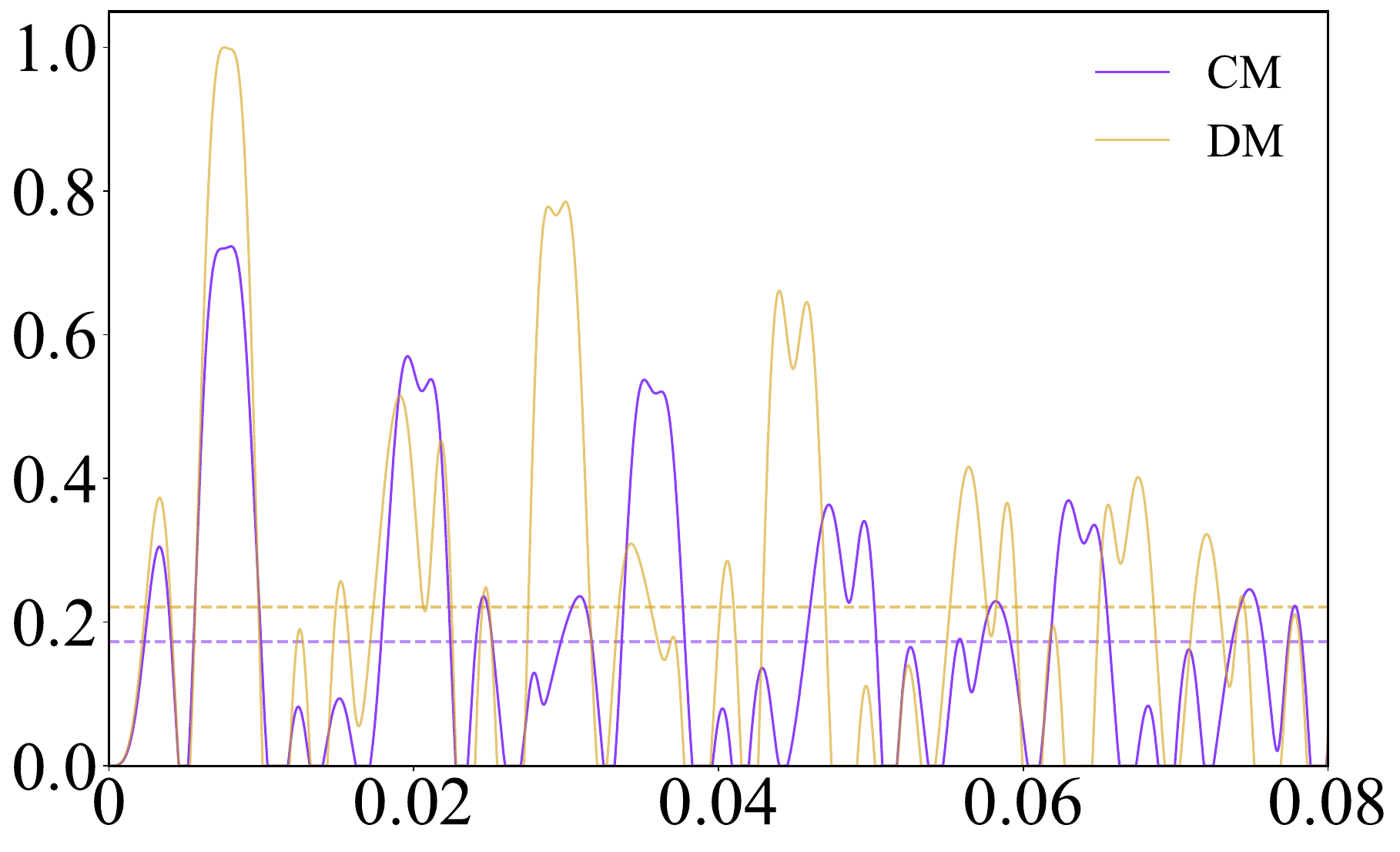}

    \makebox[0.33\textwidth][l]{\hspace{8mm}\doubleletter[0.92]{D}}%
    \makebox[0.333\textwidth][l]{\hspace{6mm}\doubleletter[0.92]{E}}%
    \makebox[0.333\textwidth][l]{\hspace{6.5mm}\doubleletter[0.92]{F}}\\
    \includegraphics[width=0.68\columnwidth]
    {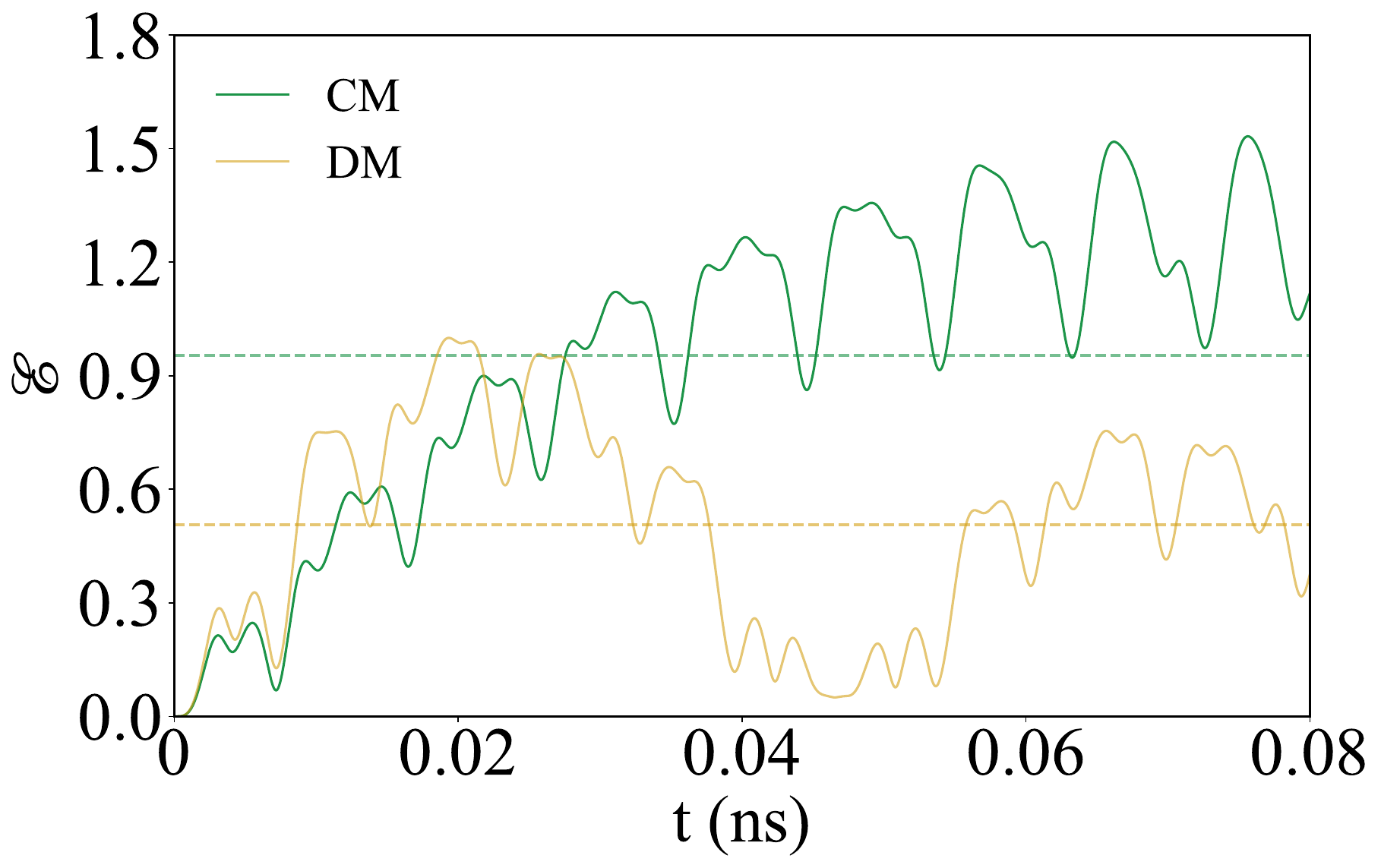}
    \hfill
    \includegraphics[width=0.68\columnwidth]
    {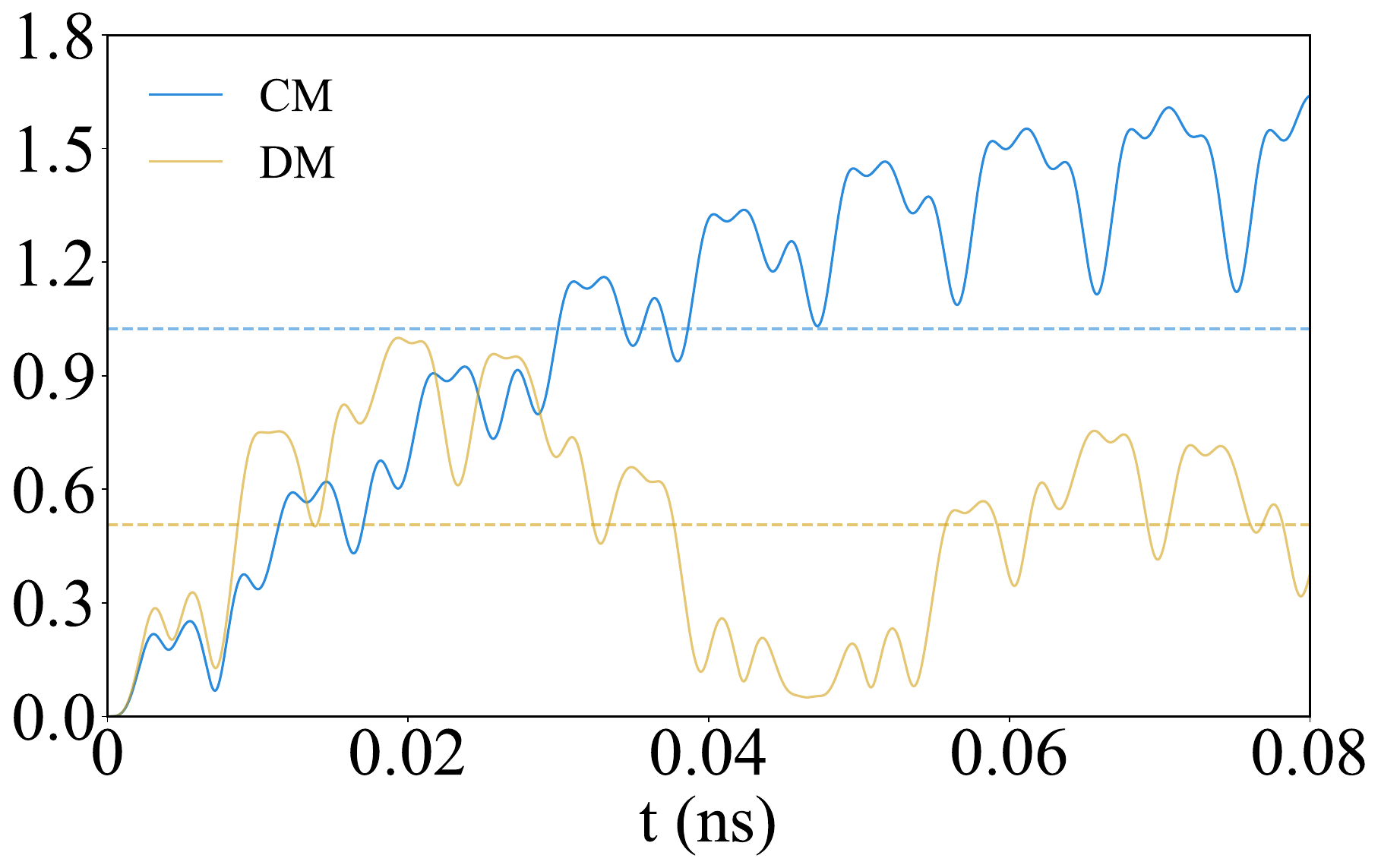}
    \hfill
    \includegraphics[width=0.67\columnwidth]{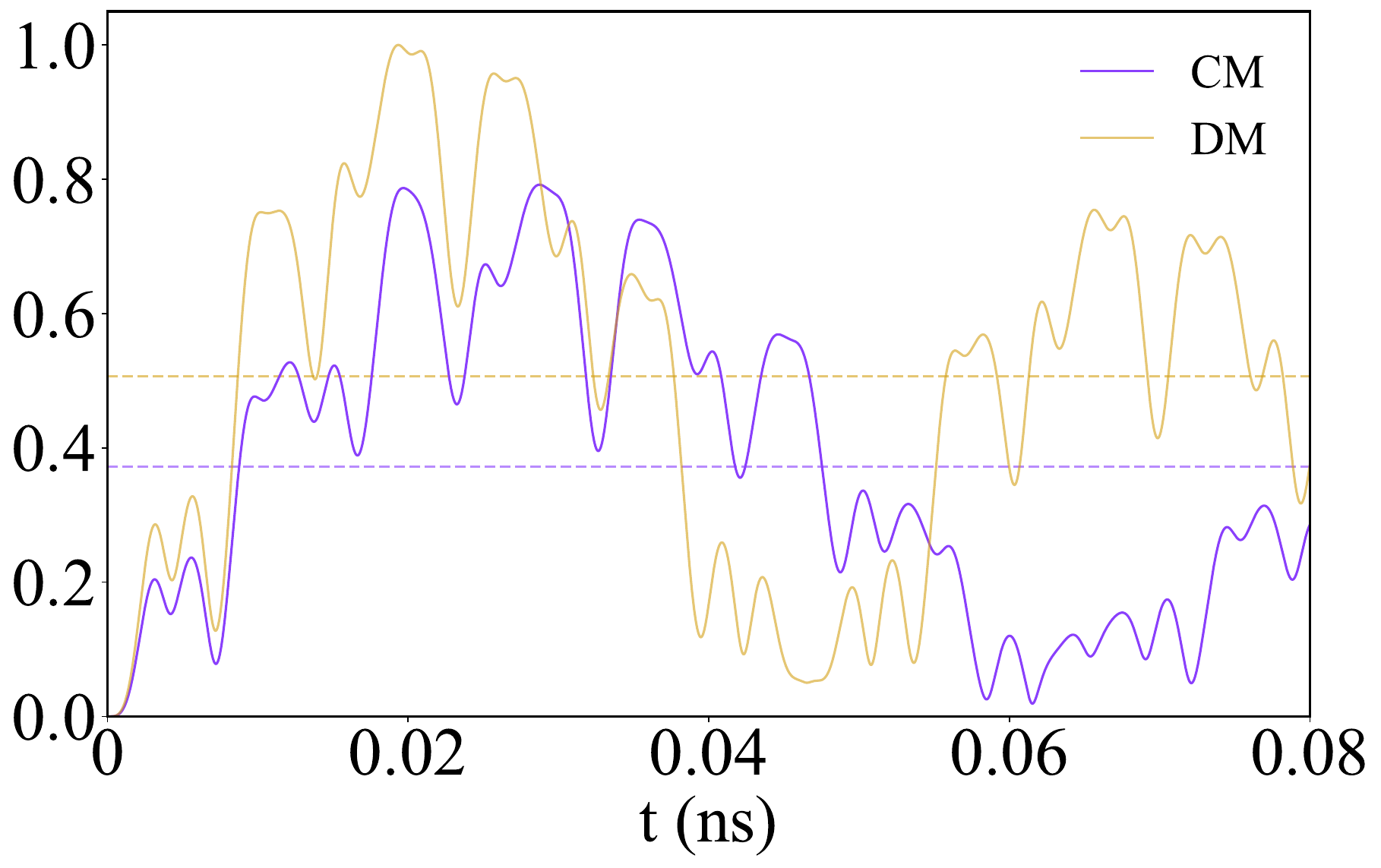}
    \captionsetup{font=small, labelfont={bf,small}}
  \caption{Comparison between the entanglement for $\text{Fe}_{8}$ obtained using the two different approaches. The columns show (from left to right) DM vs. zeroth-order CM, DM vs. first-order CM, and DM vs. second-order CM. The rows correspond to different axial anisotropy constants: first row $D$, second row $3D$. For the simulations: $B = 0.01\ \mathrm{T} \text{,} \ E = 6.02\cdot 10^9 ~\mathrm{Hz} \text{,} \ D = 3.6 \cdot 10^{10} ~\mathrm{Hz} \text{,} \ S = 3 \text{,}\ \omega = 6.75\cdot10^{11} ~\mathrm{Hz} \text{,} \ \kappa_s = 10^9~\mathrm{Hz} \text{,} \ \kappa = 7.5 \cdot 10^9~\mathrm{Hz} \ \ \text{and} \ P = 0.01 \ \mathrm{pW}$. The dashed lines indicate the entanglement averaged over the considered time interval. }
  \label{Fig:3}
\end{figure*}
\subsection{Density matrix approach}
The results obtained in the previous section, using the formalism of the covariance matrix, suggest an intrinsic quantum nature of an SMM. However, in order to quantify the entanglement, the terms in the Hamiltonian have been subjected to several approximations and assumptions.
While this was done so as to gather a transparent intuition of the phenomenology at hand, the latter should be validated with a calculation that accounts for the spin-like nature of the system. Therefore, as a last step of our study, we introduce a suitable master equation to quantify the inter-mode entanglement and compare it with the results acquired with the harmonic approach. 

To write a phenomenological master equation we recall the form of the Hamiltonian presented in Eq.~\eqref{Eq:3}, from which we remove the bath terms and maintain the description of the giant-spin, the cavity modes, the driving field and the interaction between the giant-spin and the cavity modes. Then, the corresponding Hamiltonian model takes the form
{{
\begin{equation}
	\begin{aligned}
		H &=\sum_{m} \omega_m {a}^\dagger_m {a}_m +\sum_{m} G ({a}^\dagger_m +{a}_m){S}_x+\omega_e {S}_z \\& -D{S}_z^2 + E({S}_x^2-{S}_y^2)+ i\sum_m  E_m({a}_m^\dagger e^{-i\Lambda_m t}-h.c.).
		\label{Eq:14}
	\end{aligned}
\end{equation}}}
We can describe the dynamics of the system through the phenomenological Lindblad-like master equation~\cite{PhysRevApplied.19.064060, Lindblad1976, PhysRevA.87.022337} 
\begin{equation}
\begin{aligned}
	\dot{\rho}&=-{i}[{H} \text{,} \ \rho]+\kappa_s{\cal L}_{S_z}[\rho]+\kappa\sum_m{\cal L}_{a_m}[\rho] \text{,}
    \label{Eq:15}
\end{aligned}
\end{equation}
where ${\cal L}_{O}[\rho]=2O\rho O^\dag-\{O^\dag O \text{,} \ \rho\}$  is a Lindblad superoperator for a generic operator $O$. 
The first term in Eq.~\eqref{Eq:15} accounts for the coherent evolution of the system, while the $\kappa$-dependent term describes photon dissipation for each mode. On the other hand, the term proportional to $\kappa_s$ describes pure dephasing of the giant-spin due to its interaction with the bath. Under the working assumption of weak coupling, the jump operators responsible for such mechanism are diagonal, affecting only coherences and leaving populations unchanged. In particular, for a spin $S$, they are proportional to $S_z$~\cite{Chiesa_2024}, leading to the ${\cal L}_{S_z}[\rho]$ contribution to Eq.~\eqref{Eq:15}. While, in general, decoherence of a giant-spin induced by a nuclear spin bath is known to produce a non-exponentially decaying behavior~\footnote{For some exceptions to this, see e.g. Ref.~\cite{Petiziol2021}.}, the Lindblad-like model herein allows to capture the salient features of the open dynamics of the system~\cite{PhysRevResearch.4.043135}.

The dimension of the Hilbert space plays an important role: it has  to be large enough for the simulations to be physically meaningful and small enough to be computationally manageable. In what follows, the simulations will be performed assuming that the total number of cavity modes is two, $\text{M}=2$, where each mode is described by a four-level system and the giant-spin has $S=3$. This choice is justified in Appendix~\ref{appendix7}. 
Figs.~\ref{Fig:3} show a comparison between the results achieved via the density matrix~(DM) and those corresponding to the use of the covariance matrix~(CM) for $\text{Fe}_8$. 
In the figures, the order of approximation for the Hamiltonian in the CM approach increases from left to right, with green representing the zeroth-order, blue the first-order, and purple the second-order. The order of magnitude of the entanglement is similar for the two methods and its shape varies in a similar way with respect to the value of the axial anisotropy constant -- $D$ for the first row, $3D$ for the second row -- on which the non-linearity depends. 
In principle, the second-order case (purple graphs) should be the most reliable among the three, as non-linear terms are taken into account. Indeed, it shows the best agreement with the DM results, especially as the value of the axial anisotropy constant increases (see Fig.~\ref{Fig:3}\!\!\textcolor{blue}{\doubleletter{F}}). This finding supports the conclusion that incorporating non-linear terms, via the linearization procedure, is an effective strategy for achieving a more realistic description. However, the zeroth- and first-order approximations also show good agreement with the DM results, making these descriptions useful as long as the value of $D$ remains sufficiently low.
\subsection{Extension to $\text{Mn}_{12}$}
\label{Extension}
\begin{figure}[t!]
  \centering
  \includegraphics[width=0.99\columnwidth]{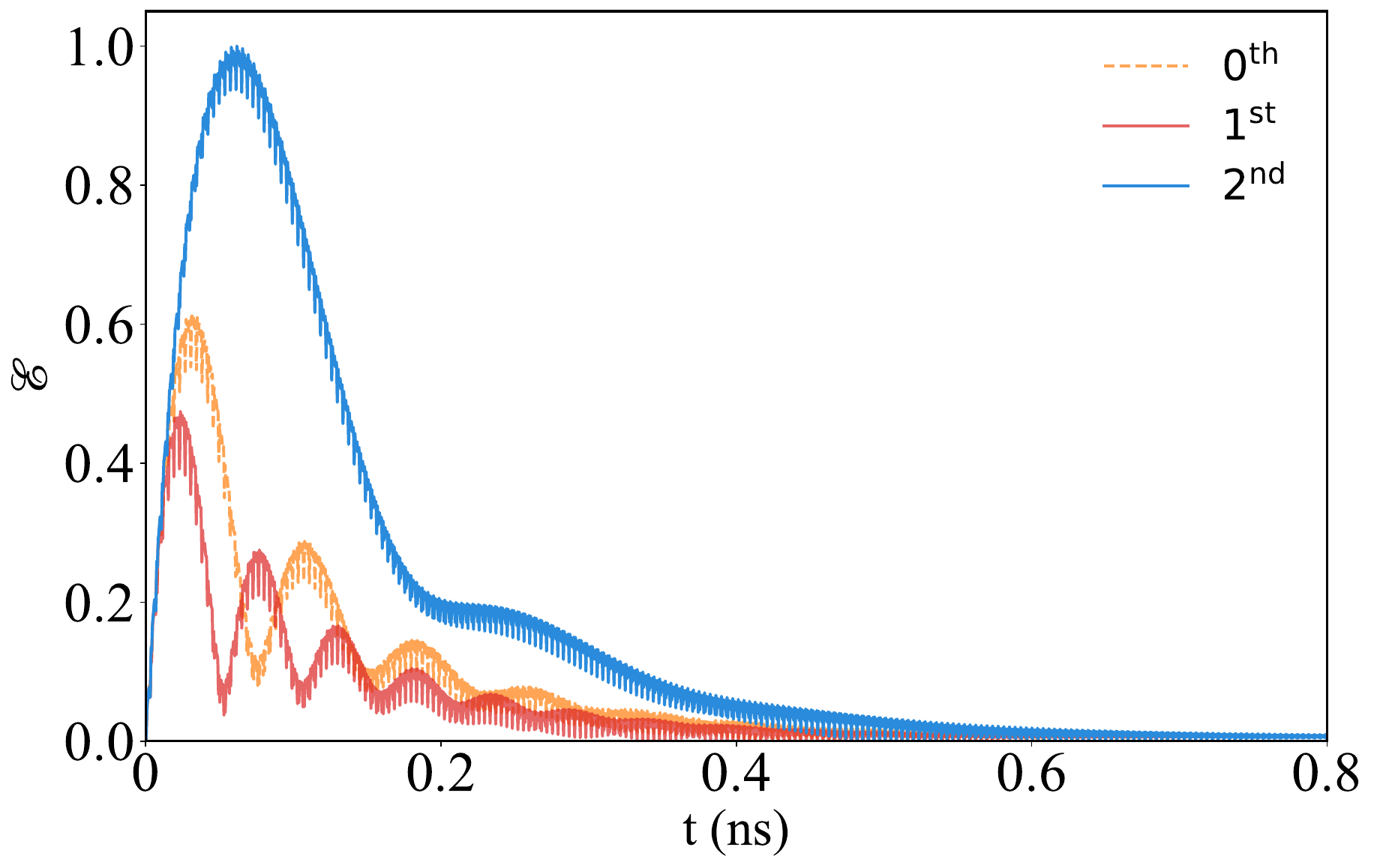}
  \captionsetup{font=small, labelfont={bf,small}}
  \caption{Bipartite entanglement for $\text{Mn}_{12}$ in the zeroth-order case (without the effect of the bath), in the first-order case (including the bath), and in the second-order case (adding the non-linear terms). The parameters used are: $ B = 0.01~\mathrm{T}\text{,} \ D=K_1= 8.64 \cdot 10^{10}~\mathrm{Hz}\text{,} \ E = 0 \text{,} \ \omega= 1.64 \cdot 10^{12}~\mathrm{Hz} \text{,} \ S = 10 \text{,} \ \alpha=\gamma=K_2=1420 \cdot 10^6~\mathrm{Hz} \ \text{and} \ J=25.$}
  \label{Fig:4}
\end{figure}

\begin{figure*}[t!]  
    \centering
    \makebox[0.33\textwidth][l]{\hspace{8mm}\doubleletter[0.92]{A}}%
    \makebox[0.333\textwidth][l]{\hspace{6mm}\doubleletter[0.92]{B}}%
    \makebox[0.333\textwidth][l]{\hspace{6.5mm}\doubleletter[0.92]{C}}\vspace{0.4mm}\\
    \includegraphics[width=0.68\columnwidth]{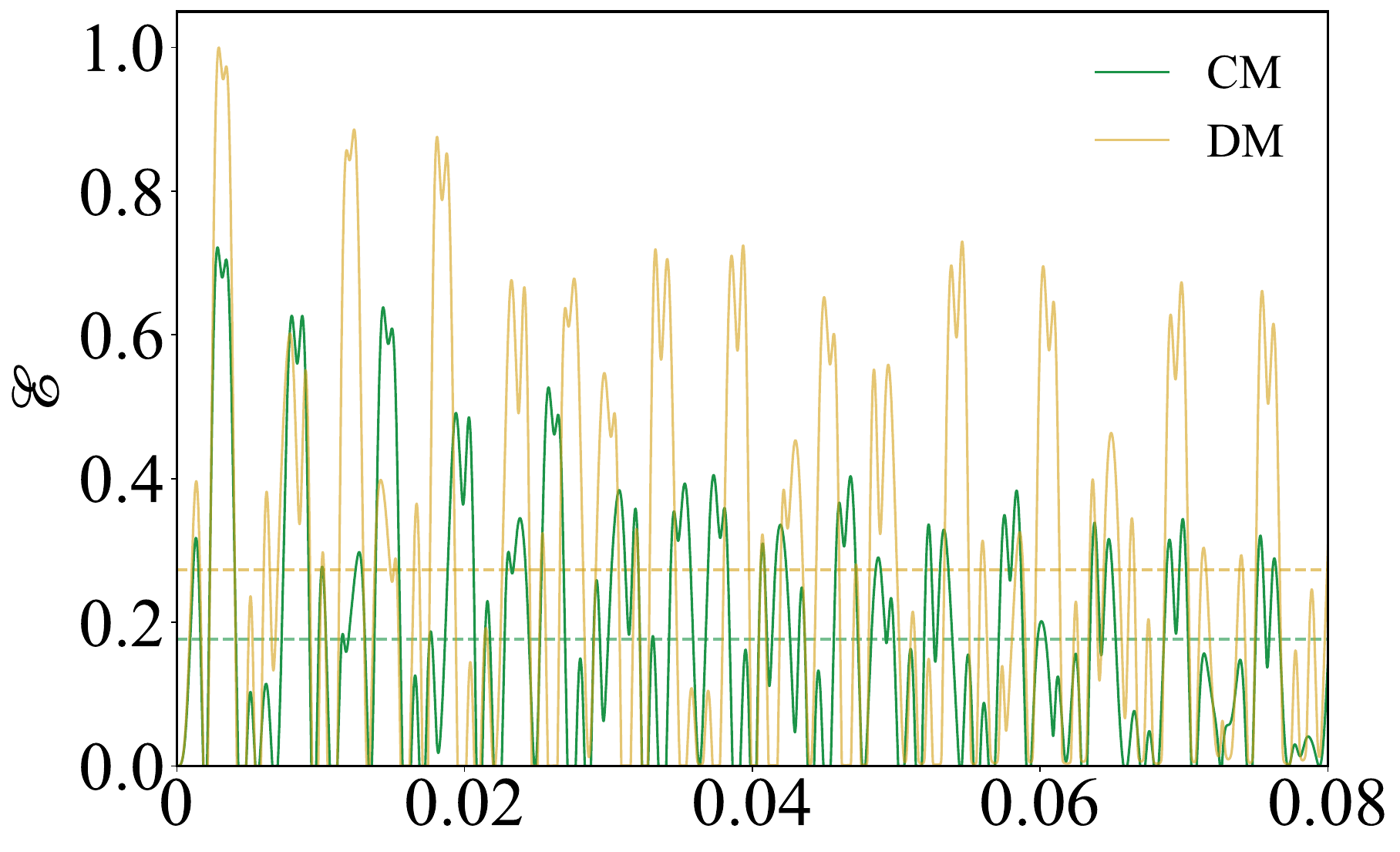}
    \hfill
    \includegraphics[width=0.68\columnwidth]{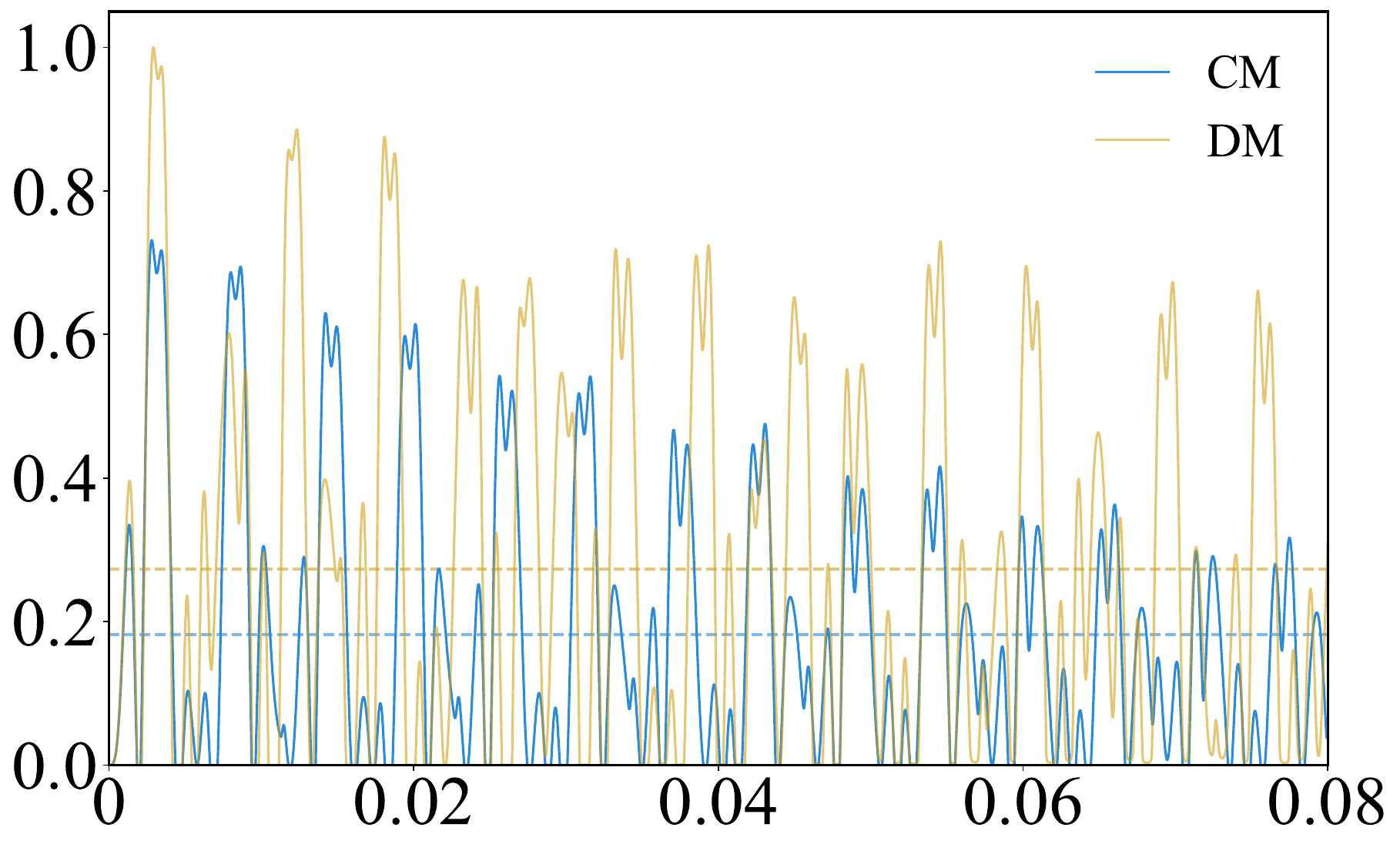}
    \hfill
    \includegraphics[width=0.68\columnwidth]
    {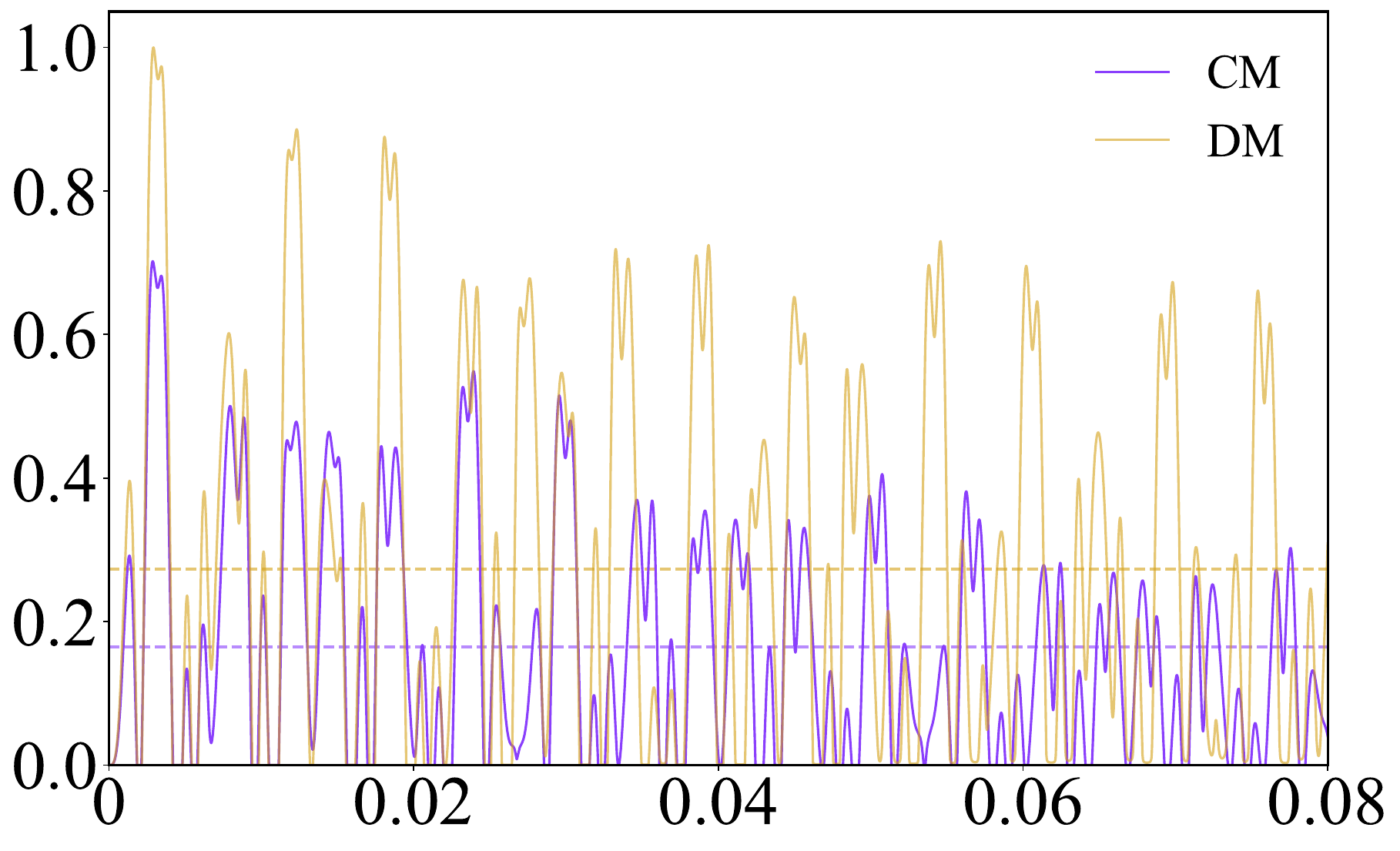}

    \makebox[0.33\textwidth][l]{\hspace{8mm}\doubleletter[0.92]{D}}%
    \makebox[0.333\textwidth][l]{\hspace{6mm}\doubleletter[0.92]{E}}%
    \makebox[0.333\textwidth][l]{\hspace{6.5mm}\doubleletter[0.92]{F}}\vspace{0.4mm}\\
    \includegraphics[width=0.68\columnwidth]
    {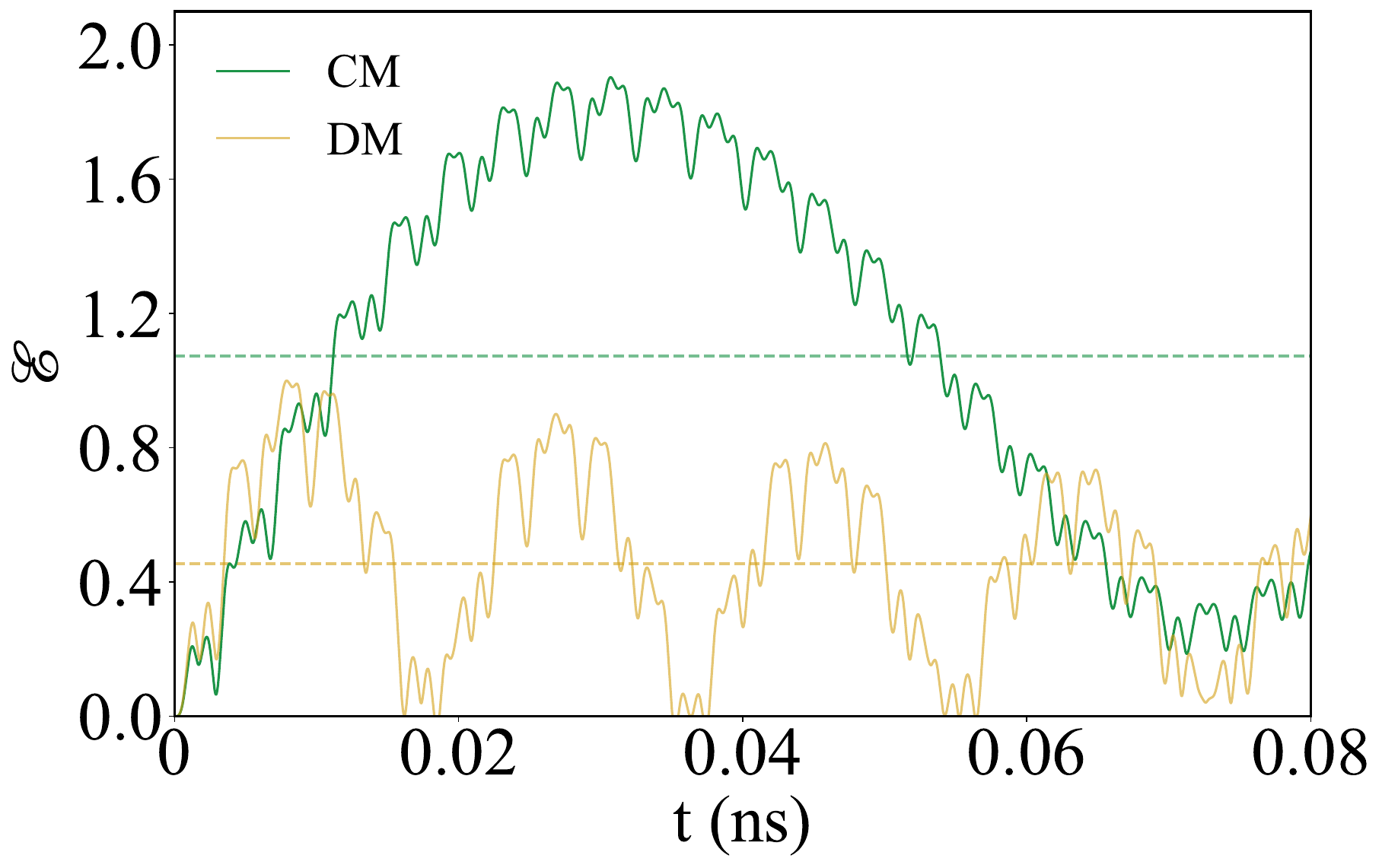}
    \hfill
    \includegraphics[width=0.68\columnwidth]
    {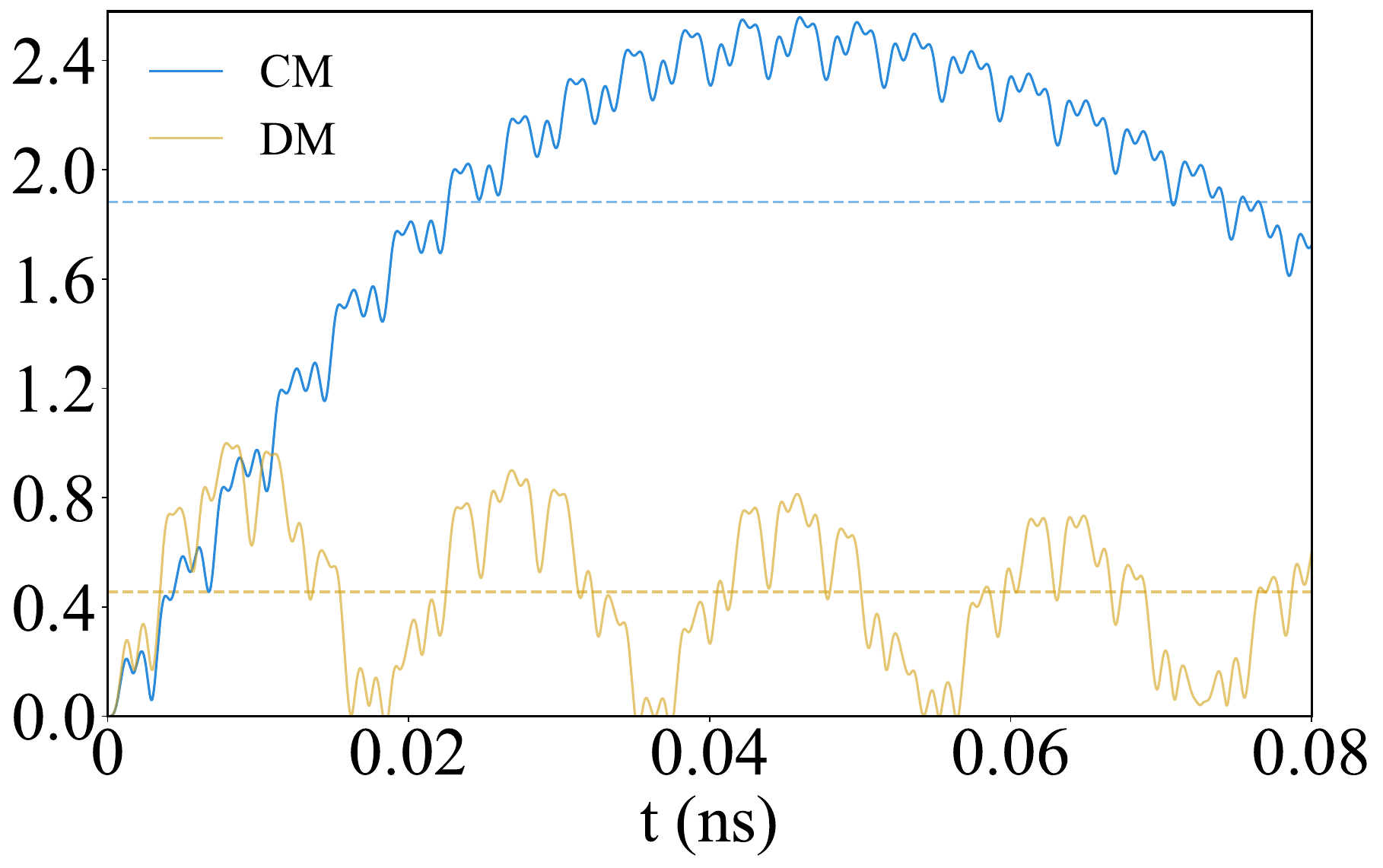}
    \hfill
    \includegraphics[width=0.68\columnwidth]{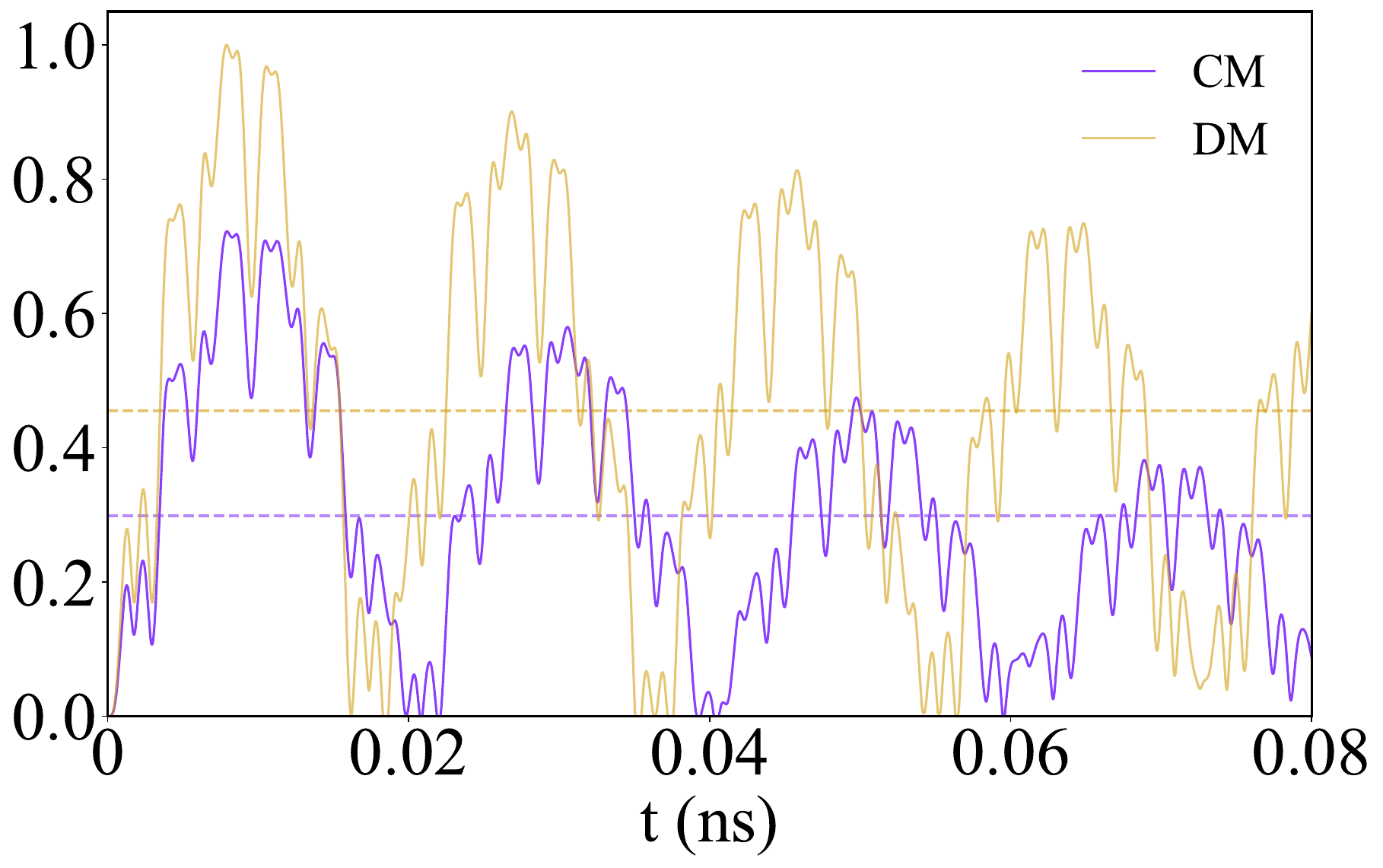}

    \captionsetup{font=small, labelfont={bf,small}}
  \caption{Comparison between the entanglement for $\text{Mn}_{12}$ obtained using the two different approaches. The columns show (from left to right) DM vs. zeroth-order CM, DM vs. first-order CM, and DM vs. second-order CM. The rows correspond to different axial anisotropy constants: first row $D$, second row $3D$. For the simulations: $B = 0.01\ \mathrm{T} \text{,}\ D  = 8.64 \cdot 10^{10} ~\mathrm{Hz} \text{,}\ E = 0 \text{,} \ S = 3 \text{,}\ \omega = 1.64\cdot10^{12} ~\mathrm{Hz}\text{,} \ \kappa_s = 10^9~\mathrm{Hz} \text{,} \ \kappa = 7.5 \cdot 10^9~\mathrm{Hz} \ \text{and} \ P = 0.01 \ \mathrm{pW}$. The dashed lines indicate the entanglement averaged over the considered time interval. }
  \label{Fig:5}
\end{figure*}

All the results presented in the main text are based on the experimental parameters of $\operatorname{Fe}_8$. However, it is also instructive to examine the behavior of a different single-molecule magnet, such as $\operatorname{Mn}_{12}$, to assess the generality of our findings. As illustrated in Figs.~\ref{Fig:4} and \ref{Fig:5}, $\operatorname{Mn}_{12}$ similarly induces entanglement between the cavity modes. This demonstrates that the creation of entanglement is not unique to $\operatorname{Fe}_8$, but rather a more general feature of such systems, highlighting their intrinsic non-classical behavior. The observations made for $\operatorname{Fe}_8$ apply in a similar manner to $\operatorname{Mn}_{12}$; however, in this case, the inclusion of nonlinear terms has a more pronounced impact on the resulting entanglement.


\section{Conclusions and outlook}
\label{conc}
We have investigated the conditions under which an SMM exhibits genuinely non-classical features. We have analyzed the way it interacts with a special form of environment, designed to probe the SMM's quantumness. We made use of the non-classicality criterion of Ref.~\cite{PhysRevLett.119.120402} which establishes that an object exhibits non-classical features if it can enhance entanglement between otherwise non-interacting probes. Starting from the Hamiltonian describing the system, we analyzed three different cases, in growing order of complexity. In all the settings, we found an increase in the value of entanglement between cavity modes, indicating a non-classical nature of the nanomagnet. To test the robustness and reliability of our results, we also proposed an alternative, parallel approach using a master equation.

To conclude this work, we stress that the overall results strongly suggest a non-classical nature of SMMs, thereby providing an answer to the initial question driving this study. Furthermore, we explored various regimes by modifying several experimentally relevant parameters: the coupling strength to the bath, the number of bath spins, the magnetic core size of the SMM ($S$), and the axial anisotropy constant ($D$). The entanglement between the modes was shown to be robust across all variations, which is crucial for the generality of our analysis. Indeed, although our investigation was based on experimental parameters of just two SMMs, $\text{Fe}_8$ and $\text{Mn}_{12}$, this robustness suggests that similar results could be expected for other SMMs.

Looking ahead, a promising direction would be to investigate a broader class of SMMs in order to better understand how specific features of their molecular structure can either enhance or hinder non-classical behaviors. In particular, expanding the study to ensembles of SMMs, rather than a single unit, will be particularly relevant, since such configurations are experimentally more feasible and can exhibit interesting collective phenomena. Furthermore, examining multi-molecule systems may reveal how intermolecular interactions and cooperative effects influence coherence and entanglement.

From a chemical perspective, further collaboration with coordination chemistry and synthetic design could target new SMM architectures with tailored anisotropies and enhanced stability. These advances would not only facilitate the exploration of non-classicality, but also potentially enable integration into practical devices, ranging from quantum memories to molecular spintronic elements.

\acknowledgements 
We acknowledge financial support from the UK funding agency EPSRC (grant EP/T028424/1), the Royal Society Wolfson Fellowship (RSWF/R3/183013), the Department for the Economy of Northern Ireland under the US-Ireland R\&D Partnership Programme, the PNRR PE Italian National Quantum Science and Technology Institute (PE0000023), and the EU Horizon Europe EIC Pathfinder project QuCoM (GA no.~10032223). SC acknowledges support
from the John Templeton Foundation Grant ID 63626.
\bibliography{biblio}
\appendix
\section{Holstein-Primakoff transformation}\label{appendix1}
The Holstein-Primakoff representation~\cite{PhysRevResearch.2.043243} of spin-S operators and spin-J operators is given by
\begin{equation}
	\begin{split}
		{S}_+=&\sqrt{2S}\sqrt{1-\frac{{s}^\dagger{s}}{2S}}{s}\text{,} \quad {S}_-=\sqrt{2S}\sqrt{1-\frac{{s}^\dagger{s}}{2S}}{s}^\dagger\text{,}\\& \qquad \qquad \qquad {S}_z=S-{s}^\dagger{s}.
		\label{Eq:A.1}
	\end{split}
\end{equation}
\begin{equation}
	\begin{split}
		J_+=&\sqrt{2J}\sqrt{1-\frac{{n}^\dagger{n}}{2J}}{n}\text{,} \quad J_-=\sqrt{2J}\sqrt{1-\frac{{n}^\dagger{n}}{2J}}{n}^\dagger\text{,}\\& \qquad \qquad \qquad  J_z={n}^\dagger{n}-J\text{,}
		\label{Eq:A.2}
	\end{split}
\end{equation}
where $J=\text{N}/2$ with $\text{N}$ equal to the number of spins in the bath. The difference in the sign of the $z$-projection operator between $S$ and $J$ stems from the sign of the Zeeman term in the Hamiltonian. Indeed, it depends on the ground state of the spins and its population.

 The expressions in Eq.~\eqref{Eq:A.1} e in Eq.~\eqref{Eq:A.2} provide an exact way to map the spin operators into bosonic operators. However, the use of these is often not convenient because the square roots are impractical to work with. One usually does a Taylor expansion around large $S \text{,} \ J \gg 1 $ and obtained the linearized version
\begin{equation}
	\begin{split}
		{S}_+=\sqrt{2S}{s}\text{,} \quad {S}_-=\sqrt{2S}{s}^\dagger\text{,} \quad {S}_z=S-{s}^\dagger{s}\text{,}
		\label{Eq:A.3}
	\end{split}
\end{equation}
\begin{equation}
	\begin{split}
		{J}_+=\sqrt{2J}{n}\text{,} \quad {J}_-=\sqrt{2J}{n}^\dagger\text{,} \quad {J}_z={n}^\dagger{n}-J.
		\label{Eq:A.4}
	\end{split}
\end{equation}
In our model, expanding around $J$ is justified by the large number of spins in the bath. A more subtle question concerns the applicability of this approach to the giant electronic spin. Nevertheless, it can be shown that the use of the linearized operators remains valid even when $S$ is not particularly large.
\section{Covariance matrix}
\label{appendix2}
We introduce the general procedure for constructing the covariance matrix starting from the Langevin equations.
From Eq.~\eqref{Eq:9}, we can derive an expression for the covariance matrix
{\small{
\begin{equation}
\begin{split}
	V(t)=W_+(t)&V(0)W_+^\text{T}(t)\\&+W_+(t)\int_0^t dt'W_-(t')DW_-^\text{T}(t')W_+^\text{T}(t')
    \label{Eq:B.1}\text{,}
\end{split}
\end{equation}}}
where $D=\text{Diag}[\kappa_1,\kappa_1,...,\kappa_M,\kappa_M, \Gamma_s,\Gamma_s ]$. We also assumed that the initial quadratures are not correlated with the noise quadratures such that the mean of the cross terms is zero. After integration, we obtain a solution for the covariance matrix in a Lyapunov-like equation
{\small{
\begin{equation}
\begin{split}
	KV(t)+&V(t)K^\text{T}=-D+KW_+(t)V(0)W_+^\text{T}(t)\\&+W_+(t)V(0)W_+^\text{T}(t)K^\text{T}+W_+(t)DW_+^\text{T}(t).
    \label{Eq:B.2}
\end{split}
\end{equation}}}
If $K $ has all negative eigenvalues, then $W_+(\infty)=0$ and the stationary covariance matrix satisfies the following relation:
\begin{equation}
    KV(\infty)+V(\infty)K^\text{T}+D=0.
    \label{Eq:B.3}
\end{equation}
Eq.~\eqref{Eq:B.2} and Eq~\eqref{Eq:B.3} can both be solved numerically to find $V$.
Then, simply by knowing the decay rates and the drift matrix of a system, we are able to construct its covariance matrix. Note that its elements are independent of the noise and pumping terms.

\section{Entanglement from covariance matrix}
\label{appendix3}
The analysis and manipulation of the second order statistical moment allows the quantification of the entanglement. We exploit the fact that every positive-definite real matrix of even dimension can be expressed in diagonal form by a symplectic transformation~\cite{Yurischev_2025}. Consequently, given a $N$-mode covariance matrix, there must exist a matrix $S$ such that
\begin{equation}
	V=S \ V^{\oplus}S^\text{T} \quad  \operatorname{where} \quad V^{\oplus}:= \displaystyle\bigoplus_{k=1}^{N}\nu_k \mathbb{I}
    \label{Eq:C.1}
\end{equation}
where the diagonal matrix $V^{\oplus}$ is called the Williamson form of $V$ and the $N$ positive quantities $\nu_k$ are called the symplectic eigenvalues of $V$~\cite{Arvind_1995}. The symplectic spectrum $\{\nu_k\}_{k=1}^N$ can be easily computed as the standard eigenspectrum of the matrix $|i\Omega V|$ where
\begin{equation}
	\Omega= \displaystyle\bigoplus_{k=1}^{N}
	\begin{pmatrix}
		0 & 1\\
		-1 & 0
	\end{pmatrix}.
    \label{Eq:C.2}
\end{equation}
For a physical covariance matrix $2\nu_k \ge 1$. 

Now, we present the operational procedure to calculate the bipartite entanglement that we use throughout the analysis. To do so, we fix the total number of modes and consider two disjoint subsets, $A$ and $B$, from the original set. Assume, now, that we want to quantify the entanglement between $A$ and $B$, $\mathscr{E}_{A:B}$. We proceed as follows:
\begin{itemize}
	\item[-] From the total covariance matrix, we only consider the entries involving the modes of subsets $A$ and $B$;
	\item[-] We apply a partial transposition to the covariance matrix with respect to the modes of $B$, $\tilde V$. From an operatorial viewpoint this is equivalent to flipping the sign of the quadrature operator ${y}$ of the $B$ modes only;
	\item[-] Finally, we check if the partially transposed covariance matrix, $\tilde V$, is physical. In fact, a non physical $\tilde V$ indicates the presence of entanglement between $A$ and $B$.
\end{itemize}
The amount of entanglement can be computed via logarithmic negativity:
	\begin{equation}
	 	\operatorname{\mathscr{E}=\max}[0\text{,}-\ln(2\tilde\nu_{min})]\text{,}
        \label{Eq:C.3}
	\end{equation}
where $\tilde\nu_{min}$ is the minimum sympletic eigenvalue of $\tilde V$.

We emphasize that, in the procedure outlined above, we obtain the partially transposed covariance matrix by flipping the sign of specific quadrature operators. This is far from being a trivial statement. It is based on the recognition that the partial transpose operation acquires, in the continuous-variable case, a geometric interpretation as mirror reflection in the Wigner phase space~\cite{PhysRevLett.84.2726}:
\begin{equation}
	{\rho} \rightarrow {\rho}^\Gamma \qquad  \Leftrightarrow \qquad W(x\text{,}y) \rightarrow W(x\text{,}-y).
    \label{Eq:C.4}
\end{equation}
\section{Zeroth-order case}
\label{appendix4}
We compute the Langevin equations starting from the Hamiltonian in Eq.~\eqref{Eq:5}
{\small
\begin{equation}
\label{Eq:D.1}
\begin{aligned}
\dot{a}_m &= -(i\omega_m + \kappa_m) a_m 
            - i G \sqrt{\frac{S}{2}} (s + s^\dagger) 
            + E_m e^{-i \Lambda_m t} 
            + \sqrt{2\kappa_m} Q_m\text{,}\\[1ex]
\dot{s}   &= -(i \Omega_s + \Gamma_s) s 
            - i G \sum_m \sqrt{\frac{S}{2}} (a_m + a_m^\dagger) 
            - 2 i E S s^\dagger 
            + \sqrt{2 \Gamma_s} F.
\end{aligned}
\end{equation}
}
We rewrite this set of equations through the quadrature operators
{\small{
\begin{equation}
\label{Eq:D.2}
\begin{aligned}
\dot{x}_m &= -\kappa_m x_m + \omega_m y_m 
            + \sqrt{2} E_m \cos(\Lambda_m t) 
            + \sqrt{2 \kappa_m} X_m\text{,}\\[1.3ex]
\dot{y}_m &= -\omega_m x_m - \kappa_m y_m 
            - G \sqrt{2S} x_s 
            - \sqrt{2} E_m \sin(\Lambda_m t) 
            \\&+ \sqrt{2 \kappa_m} Y_m\text{,}\\[1.3ex]
\dot{x}_s &= -\Gamma_s x_s + (\Omega_s - 2 E S) y_s 
            + \sqrt{2 \Gamma_s} X_s\text{,}\\[1.3ex]
\dot{y}_s &= -(\Omega_s + 2 E S) x_s - \Gamma_s y_s 
            - \sqrt{2S} G \sum_m x_m 
            + \sqrt{2 \Gamma_s} Y_s.
\end{aligned}
\end{equation}}}
For convenience the noise terms have been decomposed as ${Q}_m=({X}_m+i{Y}_m)/{\sqrt{2}}\text{,}\ {F}=({X}_s+i{Y}_s)/{\sqrt{2}}$. The dynamics of the modes can be compactly expressed as follows:
\begin{equation}
\dot{u}(t)=Ku(t)+l(t)\text{,}
\label{Eq:D.3}
\end{equation}
where $u(t)=({x}_1\text{,} \ {y}_1\text{,} \ {x}_2\text{,} \ {y}_2\text{,}...\text{,} \ {x}_M\text{,} \ {y}_M\text{,} \ {x}_s \text{,} \ {y}_s)^\text{T}$.
The drift matrix is defined as
\begin{equation}
K=
\begin{pmatrix}
	I_1 & 0 & ...& 0 & A \\
	0 & I_2 & ...& 0 & A \\
	\vdots & \vdots & \ddots& \vdots & \vdots  \\
	0 & 0 & ...& I_M& A  \\
	A & A & ...& A & S\\
\end{pmatrix}\text{,}
\end{equation}
where

\begin{equation*}
\begin{split}
I_m=&
\begin{pmatrix}
	-\kappa_m & \omega_m\\
	-\omega_m &-\kappa_m
\end{pmatrix}\text{,}
\qquad A=
\begin{pmatrix}
	0 & 0\\
	-\sqrt{2S}G & 0
\end{pmatrix}\text{,}
\\& \qquad  S=
\begin{pmatrix}
	-\Gamma_s & \Omega_s-2ES\\
	-\Omega_s-2ES& -\Gamma_s
\end{pmatrix}.
\end{split}
\end{equation*}
Last term in Eq.~\eqref{Eq:D.3} can be divided in a noise term and in a pumping term, $l(t)=\eta(t)+p(t)$ where
{\small{
\begin{equation}
\frac{\eta(t)}{\sqrt{2}}=
\begin{pmatrix}
	\sqrt{\kappa_1}{X}_1 \\
	\sqrt{\kappa_1}{Y}_1\\
	\vdots\\
	\sqrt{\kappa_m}{X}_M  \\
	\sqrt{\kappa_m}{Y}_M\\
	\sqrt{\Gamma_s}{X}_S \\
	\sqrt{\Gamma_s}{Y}_S\\
\end{pmatrix}\text{,}
\quad
\frac{p(t)}{\sqrt{2}}=
\begin{pmatrix}
	E_m\cos(\Lambda_1t) \\
	-E_m\sin(\Lambda_1t)\\
	\vdots\\
	E_m\cos(\Lambda_Mt) \\
	-E_m\sin(\Lambda_Mt)\\
	0 \\
	0\\
\end{pmatrix}.
\end{equation}}}
\section{First-order case}
\label{appendix5}
We compute the Langevin equation starting from the Hamiltonian in Eq.~\eqref{Eq:12} and rewrite them through the quadrature operators
\begin{equation}
\label{Eq:B.5}
\begin{aligned}
\dot{x}_m &= -\kappa_m x_m + \omega_m y_m 
            + \sqrt{2} E_m \cos(\Lambda_m t) 
            + \sqrt{2 \kappa_m} X_m \text{,}\\[1.4ex]
\dot{y}_m &= -\omega_m x_m - \kappa_m y_m 
            - G \sqrt{2S} x_s 
            - \sqrt{2} E_m \sin(\Lambda_m t) 
            \\& + \sqrt{2 \kappa_m} Y_m\text{,}\\[1.4ex]
\dot{x}_s &= -\Gamma_s x_s + (\Omega_s - 2 E S) y_s 
            + \alpha \sqrt{J S} y_n 
            \\& + \sqrt{2 \Gamma_s} X_s\text{,}\\[1.4ex]
\dot{y}_s &= -(\Omega_s + 2 E S) x_s - \Gamma_s y_s 
            - \sqrt{2S} G \sum_m x_m 
            \\& - \alpha \sqrt{J S} x_n 
            + \sqrt{2 \Gamma_s} Y_s\text{,}\\[1.4ex]
\dot{x}_n &= -\Gamma_b x_n + \Omega_n y_n 
            + \alpha \sqrt{J S} y_s 
            + \sqrt{2 \Gamma_b} X_b\text{,}\\[1.4ex]
\dot{y}_n &= -\Omega_n x_n - \Gamma_b y_n 
            - \alpha \sqrt{J S} x_s 
            + \sqrt{2 \Gamma_b} Y_b.
\end{aligned}
\end{equation}

The dynamics of the modes can be compactly expressed as shown in Eq.~\eqref{Eq:D.3} where, in this case, the drift matrix is defined as
\begin{equation}
	K=
	\begin{pmatrix}
		I_1 & 0 & ...& 0 & A & 0 \\
		0 & I_2 & ...& 0 & A & 0 \\
		\vdots & \vdots & \ddots& \vdots & \vdots & \vdots  \\
		0 & 0 & ...& I_M& A & 0 \\
		A & A & ...& A & S & SB\\
		0 & 0 & ...& 0 & SB & B
	\end{pmatrix}\text{,}
\end{equation}
with
\begin{equation*}
	I_m=
	\begin{pmatrix}
		-\kappa_m & \omega_m\\
		-\omega_m &-\kappa_m
	\end{pmatrix}\text{,}
	\quad A=
	\begin{pmatrix}
		0 & 0\\
		-\sqrt{2S}G & 0
	\end{pmatrix}\text{,}
\end{equation*}
\begin{equation*}
    SB=
	\begin{pmatrix}
		0 & \alpha \sqrt{JS}\\
		-\alpha \sqrt{JS} & 0
	\end{pmatrix}\text{,}
    \quad B=
	\begin{pmatrix}
		-\Gamma_b & \Omega_n\\
		-\Omega_n & -\Gamma_b
	\end{pmatrix}.
\end{equation*}
\begin{equation*}
	S=
	\begin{pmatrix}
		-\Gamma_s & \Omega_s-2ES\\
		-\Omega_s-2ES& -\Gamma_s
	\end{pmatrix}.
\end{equation*}
\\The last term in Eq.~\eqref{Eq:D.3} can be split in a noise term and in a pumping term
{\small{
\begin{equation}
	\frac{\eta(t)}{\sqrt{2}}=
	\begin{pmatrix}
		\sqrt{\kappa_1}{X}_1\\
		\sqrt{\kappa_1}{Y}_1\\
		\vdots\\
		\sqrt{\kappa_m}{X}_M \\
		\sqrt{\kappa_m}{Y}_M\\
		\sqrt{\Gamma_s}{X}_s\\
		\sqrt{\Gamma_s}{Y}_s\\
		\sqrt{\Gamma_b}{X}_b \\
		\sqrt{\Gamma_b}{Y}_b
	\end{pmatrix}\text{,}
	\qquad
	\frac{p(t)}{\sqrt{2}}=
	\begin{pmatrix}
		E_1\cos(\Lambda_1t) \\
		-E_1\sin(\Lambda_1t)\\
		\vdots\\
		E_M\cos(\Lambda_Mt) \\
		-E_M\sin(\Lambda_Mt)\\
		0 \\
		0\\
		0 \\
		0
	\end{pmatrix}.
\end{equation}}}
\section{Second-order case}
\label{appendix6}
We compute the non-linear Langevin equations starting from the Hamiltonian in Eq.~\eqref{Eq:13}
{\small{
\begin{equation}
\label{Eq:F1}
\begin{aligned}
\dot{a}&_m = -(i \omega_m + \kappa_m) a_m 
             - i G \sqrt{\frac{S}{2}} (s + s^\dagger) 
             + E_m e^{-i \Lambda_m t} 
            + \sqrt{2 \kappa_m} Q_m\text{,}\\[1.7ex]
\dot{s}   &= -\big[i (\Omega_s - K_1) + \Gamma_s \big] s 
             - i G \sqrt{\frac{S}{2}} \sum_m (a_m + a_m^\dagger) 
             \\&+ 2 i K_1 s^\dagger s s 
              - i \alpha \sqrt{J S} n 
             + i K_2 s n^\dagger n 
             - i 2 E S s^\dagger 
             + \sqrt{2 \Gamma_s} F\text{,}\\[2.5ex]
\dot{n}   &= -(i \Omega_n + \Gamma_b) n 
             - i \alpha \sqrt{J S} s 
             + i K_2 s^\dagger s n 
             + \sqrt{2 \Gamma_b} K.
\end{aligned}
\end{equation}}}
We substitute the linearized version of the operators and neglect higher-order fluctuation terms
{\small{
\begin{equation}
\label{Eq:F2}
\begin{aligned}
\langle \dot{a}_m \rangle + \delta & \dot{a}_m 
    = -(i \omega_m + \kappa_m) (\langle a_m \rangle + \delta a_m) 
       - i G \sqrt{\frac{S}{2}} \Big(\langle s \rangle + \delta s \\&+ \langle s \rangle^* + \delta s^\dagger\Big) + E_m e^{-i \Lambda_m t} 
       + \sqrt{2 \kappa_m} Q_m \text{,}\\[1ex]
\langle \dot{s} \rangle + \delta \dot{s} 
    &= -\big[i (\Omega_s - K_1) + \Gamma_s \big] (\langle s \rangle + \delta s) 
       - i \alpha \sqrt{J S} (\langle n \rangle + \delta n) \\
    &\quad - i G \sqrt{\frac{S}{2}} \sum_m (\langle a_m \rangle + \delta a_m + \langle a_m \rangle^* + \delta a_m^\dagger) \\
    &\quad + 2 i K_1 \big(|\langle s \rangle|^2 \langle s \rangle + 2 |\langle s \rangle|^2 \delta s + \langle s \rangle^2 \delta s^\dagger \big) \\
    &\quad + i K_2 \Big( |\langle n \rangle|^2 \langle s \rangle + \langle n \rangle^* \langle s \rangle \delta n + \langle n \rangle \langle s \rangle \delta n^\dagger + |\langle n \rangle|^2 \delta s \Big) \\
    &\quad - i 2 E S (\langle s \rangle^* + \delta s^\dagger) 
       + \sqrt{2 \Gamma_s} F\text{,}\\[1ex]
\langle \dot{n} \rangle + \delta \dot{n} 
    &= -(i \Omega_n + \Gamma_b) (\langle n \rangle + \delta n) 
       - i \alpha \sqrt{J S} (\langle s \rangle + \delta s) \\
    &\quad + i K_2 \Big( |\langle s \rangle|^2 \langle n \rangle + |\langle s \rangle|^2 \delta n + \langle s \rangle^* \langle n \rangle \delta s + \langle s \rangle \langle n \rangle \delta s^\dagger \Big) \\
    &\quad + \sqrt{2 \Gamma_b} K.
\end{aligned}
\end{equation}}}
Due to the complicated Hamiltonian, linearizing the Langevin equations proves to be a nontrivial task. The first step is to compute the mean amplitudes of the different modes, namely $\langle s\rangle \text{,} \langle n \rangle \text{,} \langle a_m \rangle$.
The Hamiltonian in Eq.~\eqref{Eq:13} is time-dependent. Generally, the explicit dependence on time can be removed by rotating the system with the drive. In this particular case, due to the form of the Hamiltonian, it is not possible to completely remove the explicit time dependence. However, there exists a frame that makes our calculations easier to handle. Hence, we apply the following transformation to the operators at hand
{\small
\begin{equation}
\begin{split}
	U &= \prod_m U^m \, U^s \, U^n\text{,}\\
	U &= \exp\Bigg[i \Big(\sum_m \Lambda_m a_m^\dagger a_m + \Lambda_1 s^\dagger s + \Lambda_1 n^\dagger n \Big) t \Bigg].
\end{split}
\label{Eq:F3}
\end{equation}}

In the new frame, after the rotating wave approximation, the giant-spin decouples from all modes except the first. The Hamiltonian takes the following form
{\small{
\begin{equation}
\begin{aligned}
		{H}_R&=\sum_m \Delta_m {a}_m^\dagger{a}_m+\Delta_s{s}^\dagger {s}+\Delta_n{n}^\dagger{n}+ G\sqrt{\frac{S}{2}}({a}_1^\dagger{s}+{a}_1{s}^\dagger)\\&- K_1({s}^\dagger{s})^2+\alpha\sqrt{JS}({s}{n}^\dagger+{s}^\dagger{n})+\sum_mi E_m({a}_m^\dagger-{a}_m)\\&- K_2{s}^\dagger{s}{n}^\dagger{n}.
        \label{Eq:F.4}
\end{aligned}
\end{equation}}}
From this Hamiltonian, we can easily evaluate the stationary mean values for the different modes exploiting that the derivative goes to zero. This system admits the following analytical solutions:
{\small{
\begin{equation}
\label{Eq:F.5}
\begin{aligned}
\langle a_1 &\rangle = \frac{-i G \sqrt{\frac{S}{2}} \langle s \rangle + E_1}{\kappa_1}\text{,} \\
\langle a_m &\rangle = \frac{E_m}{\kappa_m}\text{,}  \quad m = \text{,}  \dots\text{,}  6 \ \text{,} \\[1ex]
\langle n \rangle &= -\frac{i \alpha \sqrt{J S} \langle s \rangle}{i (\Delta_n - K_2 |\langle s \rangle|^2) + \Gamma_b}\text{,} \\[1ex]
\langle s \rangle &= -\frac{i G \sqrt{\frac{S}{2}} \frac{E_1}{\kappa_1}}{%
i \Big( \Delta_s - K_1 - 2 K_1 |\langle s \rangle|^2 
    - \frac{\alpha^2 J S K_2 |\langle s \rangle|^2}{\Gamma_b^2 + (\Delta_n - K_2 |\langle s \rangle|^2)^2} \Big) + B}.
\end{aligned}
\end{equation}}}
with:
\begin{equation}
    B=\Gamma_s+\dfrac{G^2S}{2k_1}+\dfrac{\alpha^2JS}{i\big(\Delta_n-K_2|\langle s \rangle|^2\big)+\Gamma_b}\text{,}
    \label{Eq:F.6}
\end{equation}
and exploiting $\Delta_m=0$, for $m=1 \text{,}...\text{,}6$.

Once we evaluate the mean value of the amplitudes, we move back to our starting frame and consider the fluctuaction part of each equation
\begin{equation}
\label{Eq:F.7}
\begin{aligned}
\delta \dot{a}&_m 
    = -(i \omega_m + \kappa_m) \delta a_m 
       - i G \sqrt{\frac{S}{2}} (\delta s + \delta s^\dagger) 
       + \sqrt{2 \kappa_m} Q_m \text{,}\\[1.5ex]
\delta \dot{s} 
    &= -\big[i (\Omega_s - K_1) + \Gamma_s \big] \delta s 
       - i \alpha \sqrt{J S}\delta n 
       \\
    &- i G \sqrt{\frac{S}{2}} \sum_m (\delta a_m + \delta a_m^\dagger)- i 2 E S \delta s^\dagger  \\
    &+ 2 i K_1 \Big( 2 |\langle s \rangle|^2 \delta s + \langle s \rangle^2 \delta s^\dagger \Big) \\
    & + i K_2 \Big( \langle n \rangle^* \langle s \rangle \delta n + \langle n \rangle \langle s \rangle \delta n^\dagger + |\langle n \rangle|^2 \delta s \Big) + \sqrt{2 \Gamma_s} F\text{,}\\[1.5ex]
\delta \dot{n} 
    &= -(i \Omega_n + \Gamma_b) \delta n 
       - i \alpha \sqrt{J S} \, \delta s 
       + i K_2 \Big( |\langle s \rangle|^2 \delta n \\&+ \langle s \rangle^* \langle n \rangle \delta s + \langle s \rangle \langle n \rangle \delta s^\dagger \Big) + \sqrt{2 \Gamma_b} K.
\end{aligned}
\end{equation}
We do not know the values of $\langle a_m \rangle$, $\langle s \rangle$, and $\langle n \rangle$ in this frame, but we have already computed them in a rotating frame. Since these are scalar quantities, their values remain unchanged by the change of frame, a part for a phase factor, so we can use the results previously recovered. We found that $\langle s \rangle$ and $\langle n \rangle$ are purely imaginary, while $\langle a_m \rangle$ is real. Using $\langle n \rangle^* = -\langle n \rangle$ and $\langle s \rangle^* = -\langle s \rangle$, we obtain the final linearized form of the Langevin equations. We then rewrite these in terms of quadrature operators as
{\small{
\begin{equation}
\label{Eq:F7}
\begin{aligned}
\dot{x}_m 
    &= -\kappa_m x_m + \omega_m y_m + \sqrt{2 \kappa_m} X_m\text{,}\\
\dot{y}_m  &= -\omega_m x_m - \kappa_m y_m - \sqrt{2S} G x_s + \sqrt{2 \kappa_m} Y_m \text{,}\\[1.5ex]
\dot{x}_s 
    &= (\Omega_{NL} + 2 K_1 \langle s \rangle^2 - 2 E S) y_s 
       - \Gamma_s x_s 
      \\& + \Big(\alpha \sqrt{J S} + 2 K_2 \langle s \rangle \langle n \rangle\Big) y_n 
       + \sqrt{2 \Gamma_s} X_s\text{,}\\
\dot{y}_s 
    &= (-\Omega_{NL} + 2 K_1 \langle s \rangle^2 - 2 E S) x_s 
       - \Gamma_s y_s 
       - G \sqrt{2 S} \sum_m x_m 
       \\&- \alpha \sqrt{J S} x_n 
       + \sqrt{2 \Gamma_s} Y_s\text{,}\\[1.5ex]
\dot{x}_n 
    &= -(\Omega_n + K_2 |\langle s \rangle|^2) y_n 
       - \Gamma_b x_n 
      \\& + (\alpha \sqrt{J S} + 2 K_2 \langle s \rangle \langle n \rangle) y_s 
       + \sqrt{2 \Gamma_b} X_b\text{,}\\[1.51ex]
\dot{y}_n 
    &= (\Omega_n + K_2 |\langle s \rangle|^2) x_n 
       - \Gamma_b y_n 
       - \alpha \sqrt{J S} x_s 
       + \sqrt{2 \Gamma_b} Y_b.
\end{aligned}
\end{equation}}}

where $\Omega_{NL}=\Omega_s-K_1-4K_1|\langle s\rangle|^2-K_2|\langle n\rangle|^2$. 

\begin{figure*}[t!]  
    \centering
    \makebox[0.33\textwidth][l]{\hspace{8mm}\doubleletter[0.92]{A}}%
    \makebox[0.333\textwidth][l]{\hspace{6mm}\doubleletter[0.92]{B}}%
    \makebox[0.333\textwidth][l]{\hspace{6.5mm}\doubleletter[0.92]{C}}\vspace{0.4mm}\\
    \includegraphics[width=0.68\columnwidth]{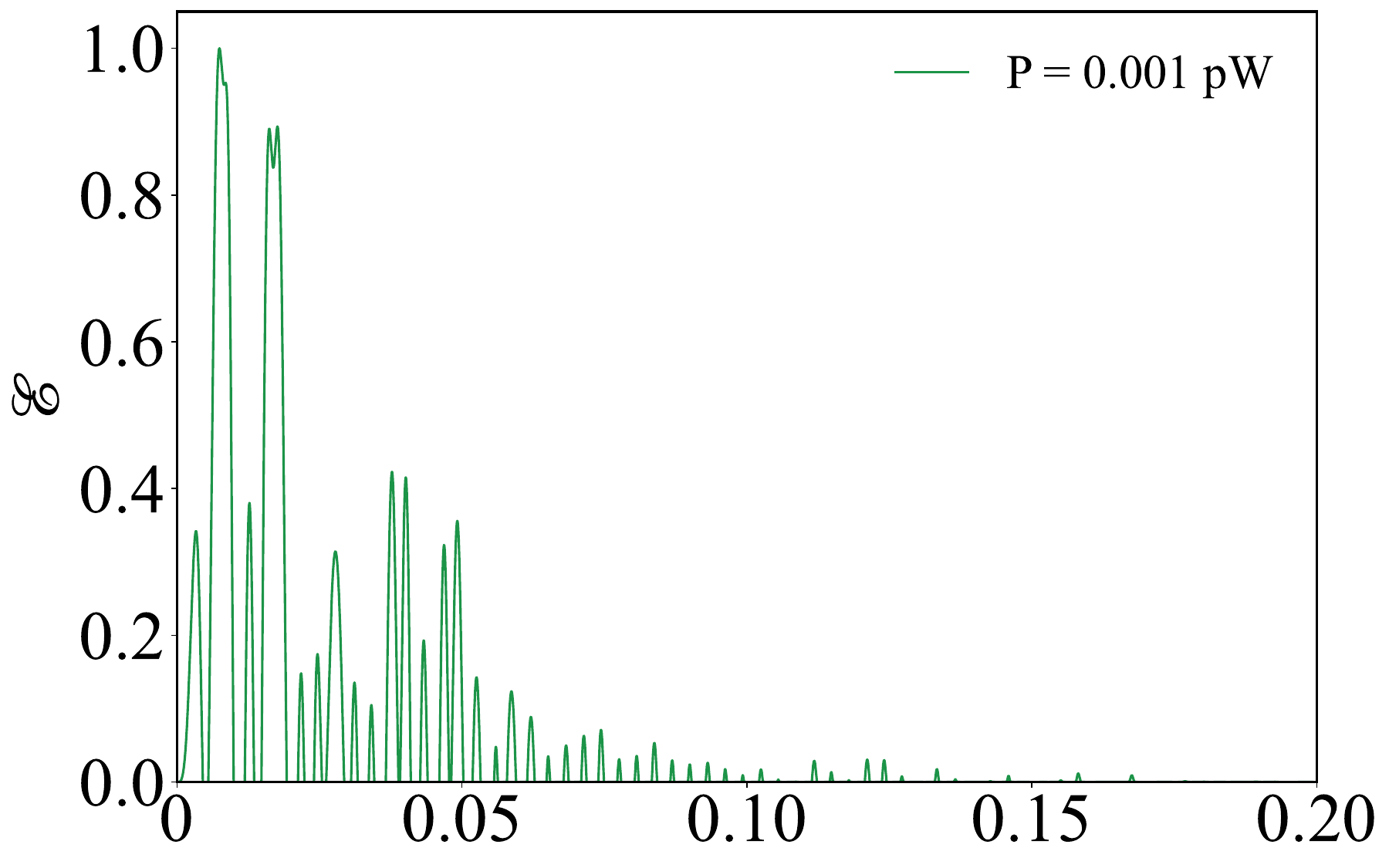}
    \hfill
    \includegraphics[width=0.68\columnwidth]{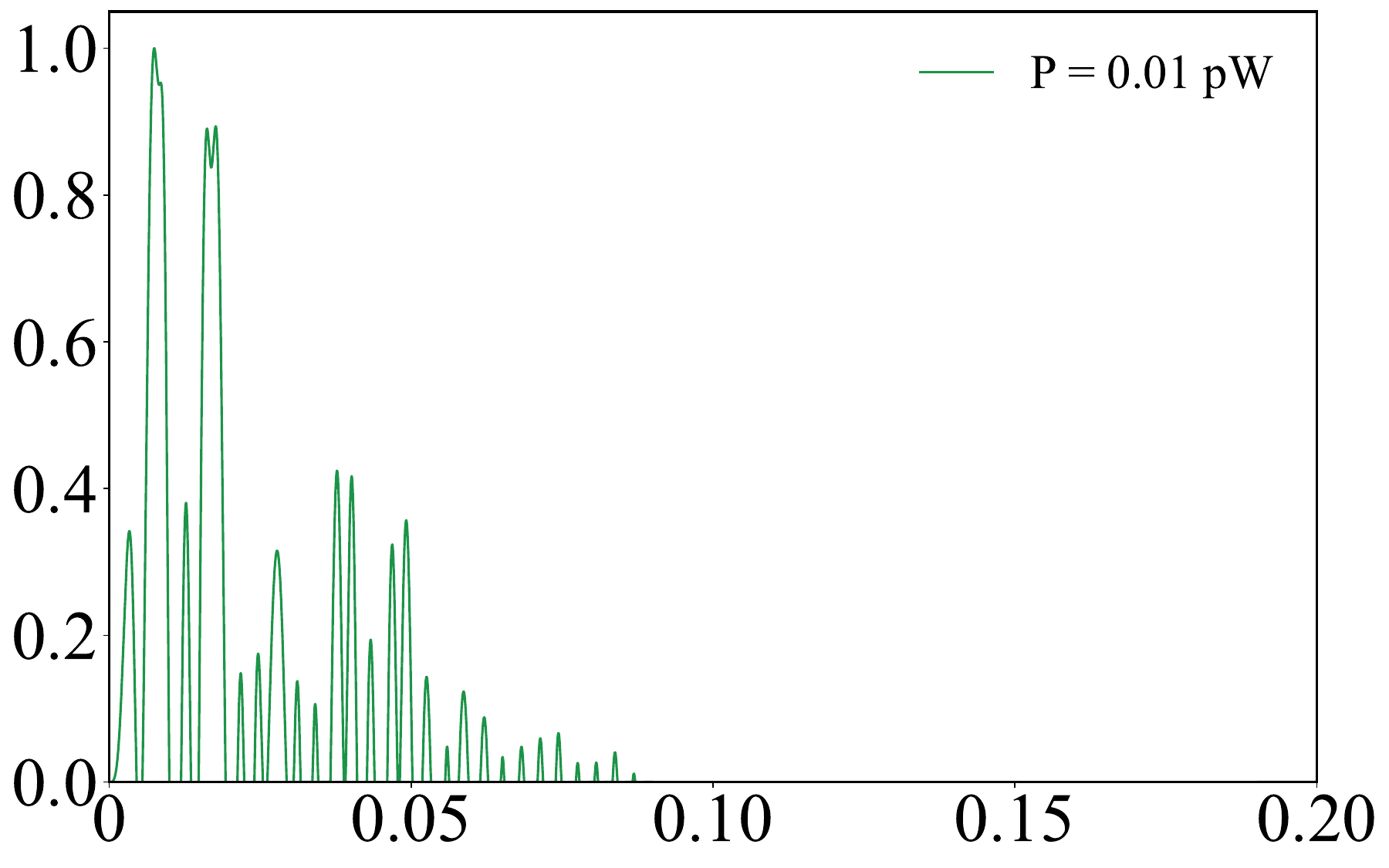}
    \hfill
    \includegraphics[width=0.68\columnwidth]
    {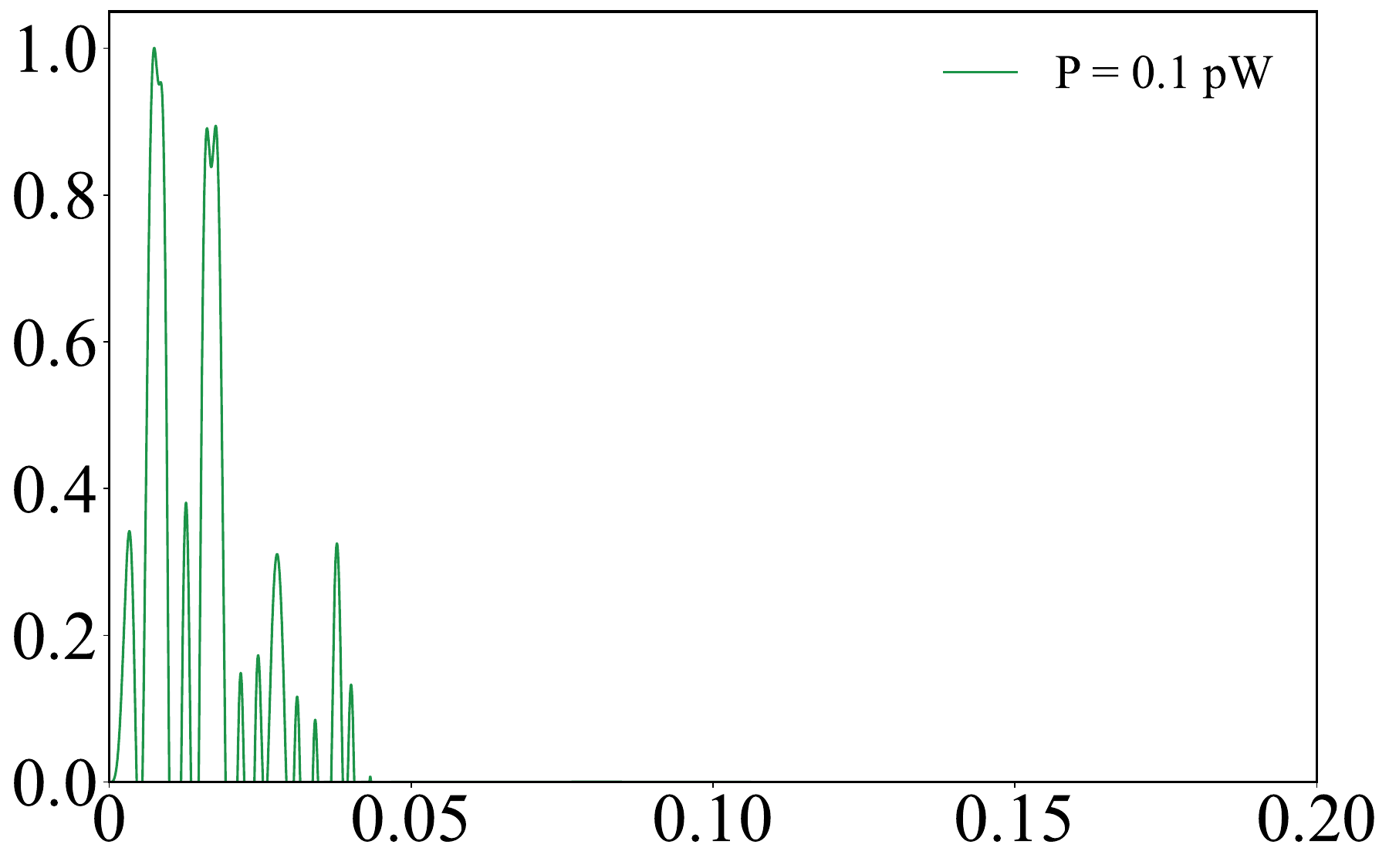}

    \makebox[0.33\textwidth][l]{\hspace{8mm}\doubleletter[0.92]{D}}%
    \makebox[0.333\textwidth][l]{\hspace{6mm}\doubleletter[0.92]{E}}%
    \makebox[0.333\textwidth][l]{\hspace{6.5mm}\doubleletter[0.92]{F}}\vspace{0.4mm}\\
    \includegraphics[width=0.68\columnwidth]
    {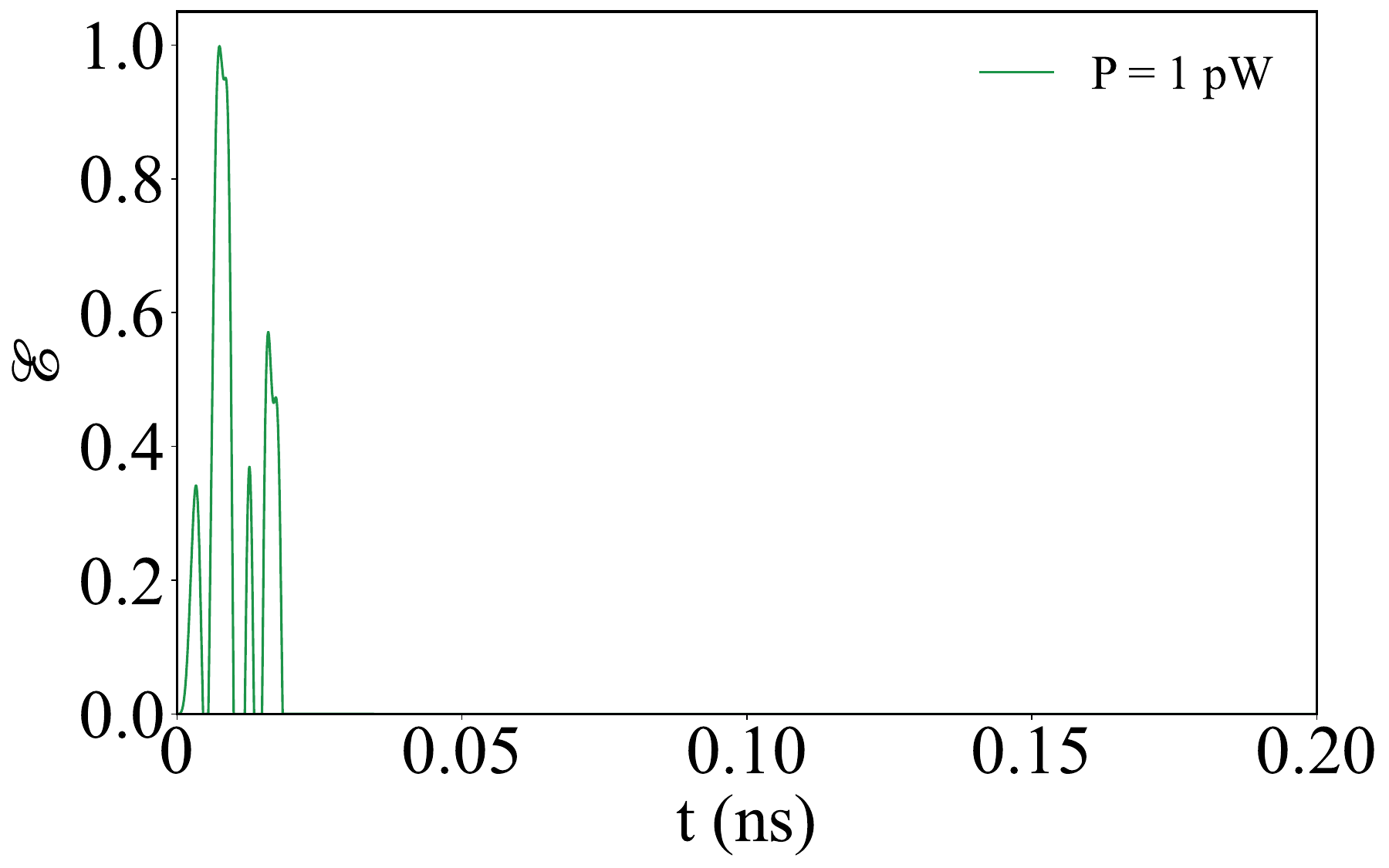}
    \hfill
    \includegraphics[width=0.68\columnwidth]
    {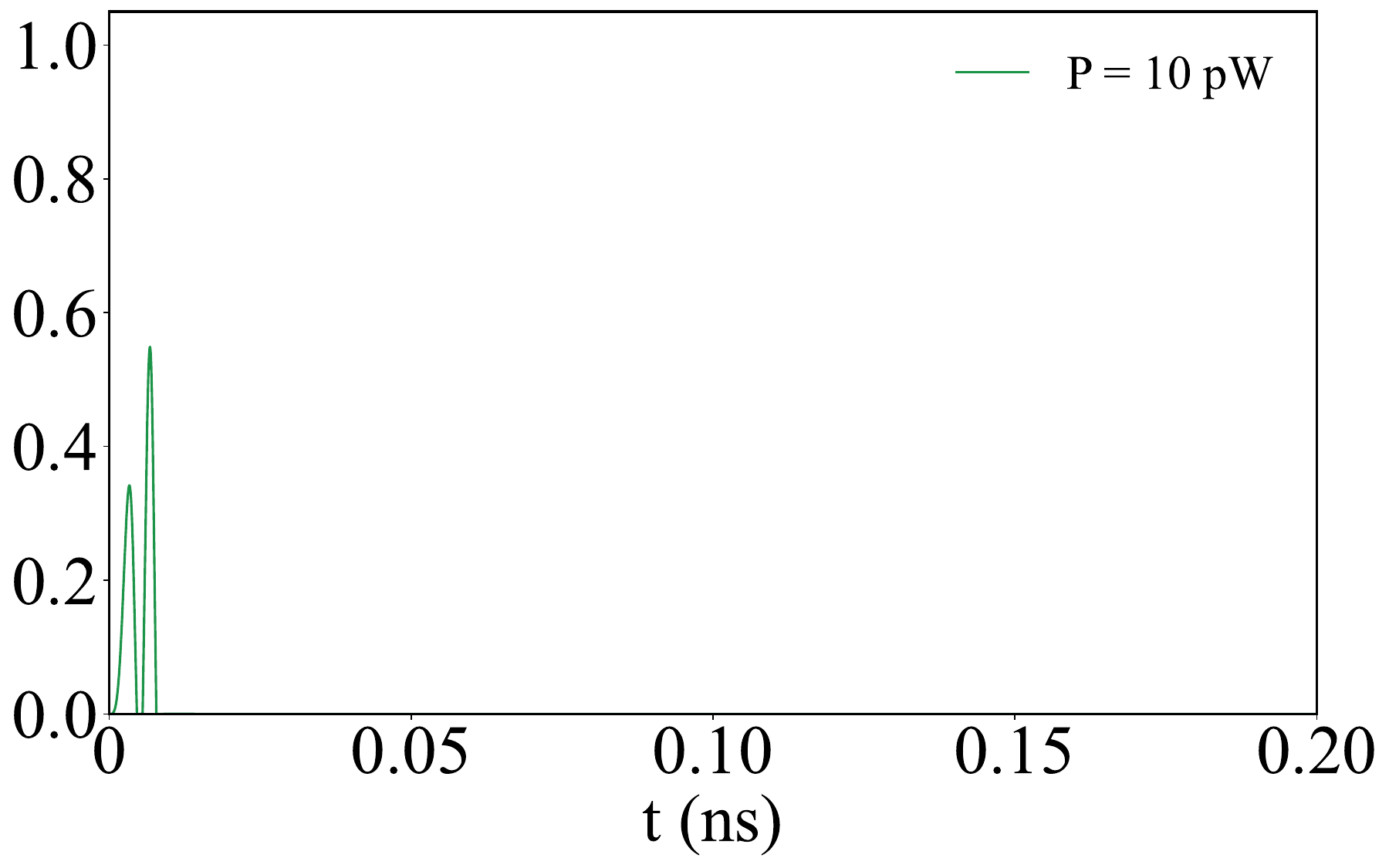}
    \hfill
    \includegraphics[width=0.68\columnwidth]{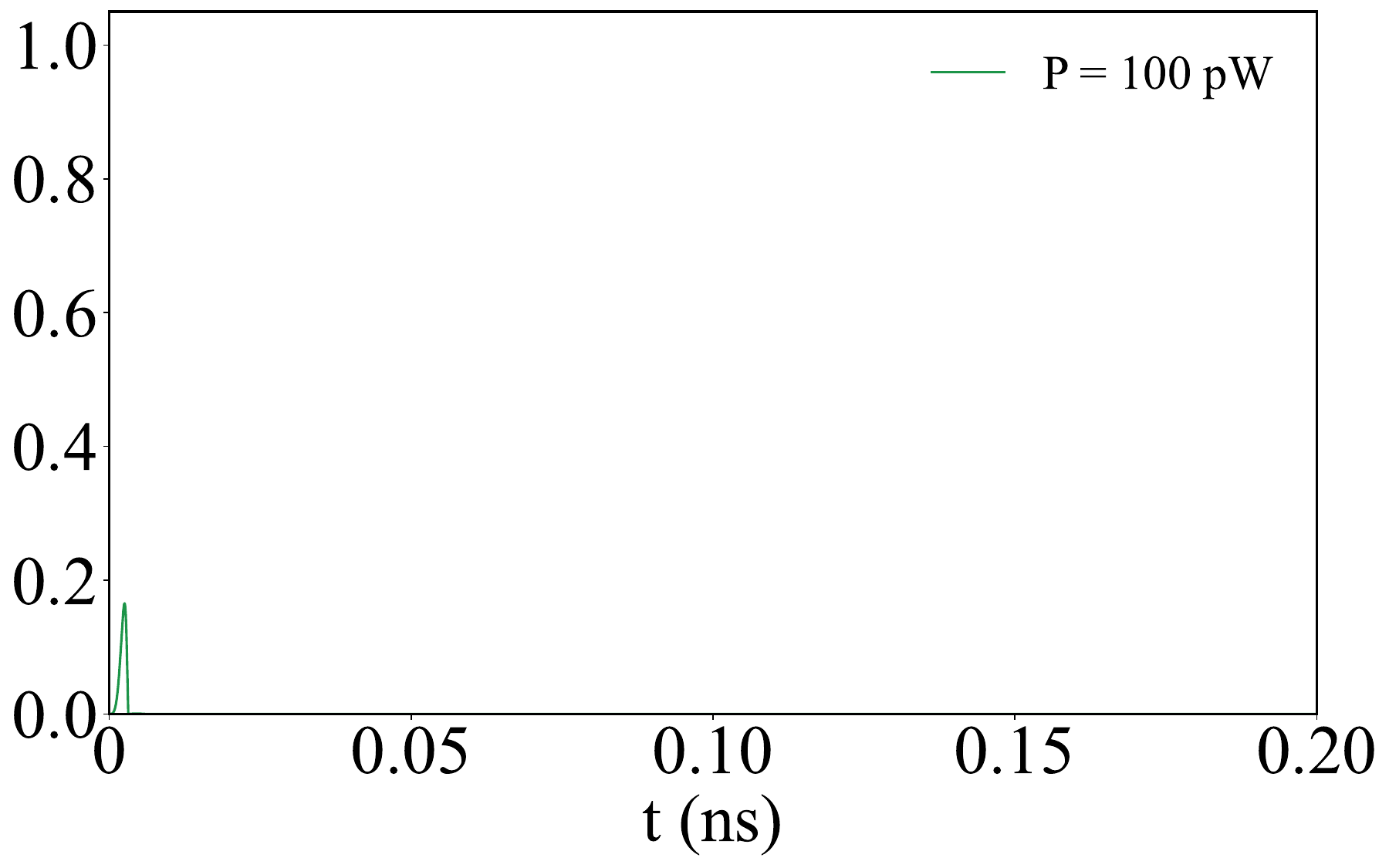}

    \captionsetup{font=small, labelfont={bf,small}}
  \caption{Bipartite entanglement between two cavity modes, interacting via a spin $S=2$, varying the value of driving power. The other parameters used for the simulations are: $B = 0.01~\mathrm{T}\text{,} \ E = 6.02\cdot 10^9~\mathrm{Hz}\text{,}  \ D = 3.6 \cdot 10^{10}~\mathrm{Hz}\text{,}  \ S = 2\text{,}  \ \omega = 6.75\cdot10^{11} ~\mathrm{Hz} \text{,}  \ \Gamma_s=10^9~\mathrm{Hz} \ \text{,} \ \kappa_s = 10^9~\mathrm{Hz} \ \text{and} \ \kappa=7.5\cdot10^9~\mathrm{Hz}. $ }
  \label{Fig:10}
\end{figure*}
We point out that the self-Kerr term modifies the frequency of the giant-spin, the cross-Kerr term modifies the bath and the giant-spin frequencies, and, in addition, introduces an interaction between both. Therefore, the matrix modifications after the introduction of the non-linear terms seem physically reasonable. 
The dynamics of the modes can be compactly expressed as shown in Eq.~\eqref{Eq:D.3} where, in this case, the drift matrix is defined as
\begin{equation}
	K=
	\begin{pmatrix}
		I_1 & 0 & ...& 0 & A & 0 \\
		0 & I_2 & ...& 0 & A & 0 \\
		\vdots & \vdots & \ddots& \vdots & \vdots & \vdots  \\
		0 & 0 & ...& I_M& A & 0 \\
		A & A & ...& A & S & SB \\
		0 & 0 & ...& 0 & SB & B
	\end{pmatrix}\text{,}
\label{Eq:F.9}
\end{equation}
with
\begin{equation*}
	{I_m}=
	\begin{pmatrix}
		-\kappa_m & \omega_m\\
		-\omega_m &-\kappa_m
	\end{pmatrix} \text{,}
	\qquad A=
	\begin{pmatrix}
		0 & 0\\
	  - \sqrt{2S}G& 0
	\end{pmatrix}\text{,}
\end{equation*}
\begin{equation*}
	SB=
	\begin{pmatrix}
		0 & \alpha \sqrt{JS}+2K_2\langle s \rangle \langle n \rangle \\
		-\alpha \sqrt{JS}& 0
	\end{pmatrix}\text{,} 
\end{equation*}
\begin{equation*}
	S=
	\begin{pmatrix}
		-\Gamma_s & \Omega_{NL} + 2K_1 \langle s\rangle^2 - 2ES\\
 		-\Omega_{NL} + 2K_1 \langle s\rangle^2 - 2ES& -\Gamma_s
	\end{pmatrix} \text{,}
\end{equation*}
\begin{equation*}
	\qquad B=
	\begin{pmatrix}
		-\Gamma_b & -\Omega_n-K_2|\langle s \rangle|^2\\
		-\Omega_n+K_2|\langle s \rangle|^2& -\Gamma_b
	\end{pmatrix}.
\end{equation*}
In this case, the last term in Eq.~\eqref{Eq:D.3} describes only the noise term, since the pumping term has been already taken into account for the calculation of the expectation values, $l(t)=\eta(t)$ with
\begin{equation}
	\frac{\eta(t)}{\sqrt{2}}=
	\begin{pmatrix}
		\sqrt{\kappa_1}{X}_1\\
		\sqrt{\kappa_1}{Y}_1\\
		\vdots\\
		\sqrt{\kappa_m}{X}_M \\
		\sqrt{\kappa_m}{Y}_M\\
		\sqrt{\Gamma_s}{X}_s\\
		\sqrt{\Gamma_s}{Y}_s\\
		\sqrt{\Gamma_b}{X}_b \\
		\sqrt{\Gamma_b}{Y}_b
	\end{pmatrix}.
\label{Eq:F.10}
\end{equation}
\begin{figure*}[t!]  
    \centering
    \hspace{5mm}\includegraphics[width=0.55\columnwidth]{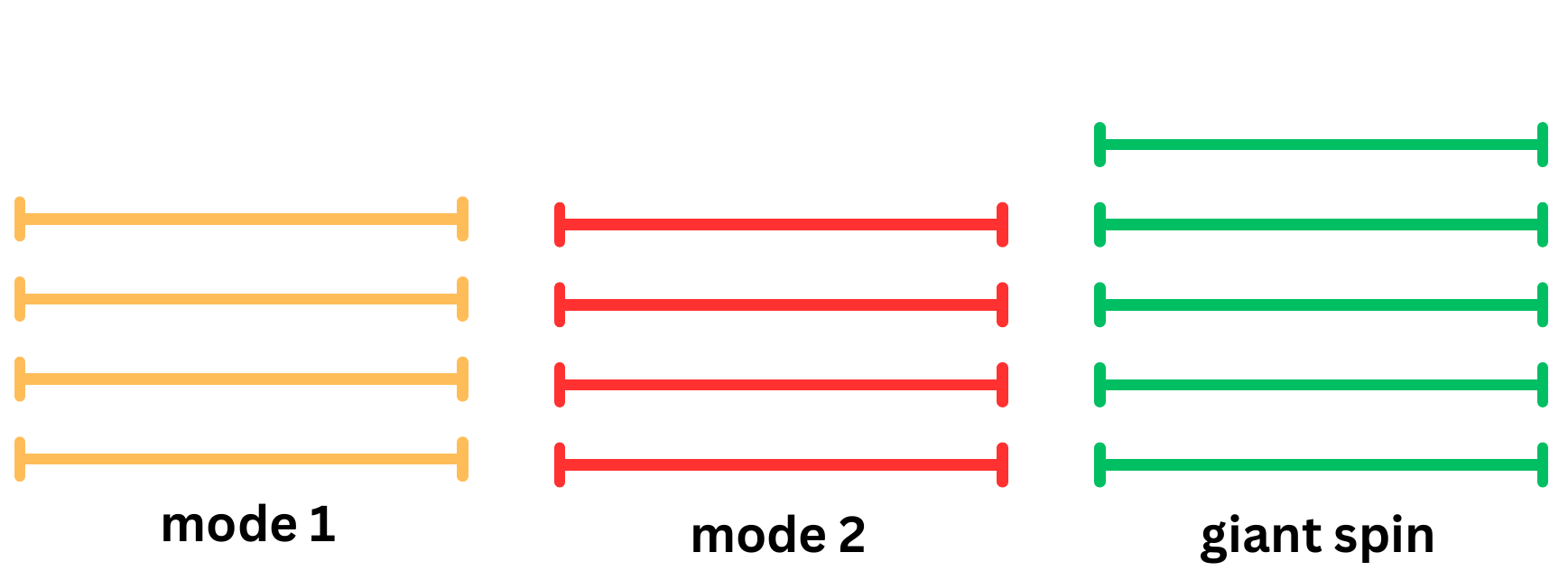}\hspace{13.5mm}
    \includegraphics[width=0.55\columnwidth]{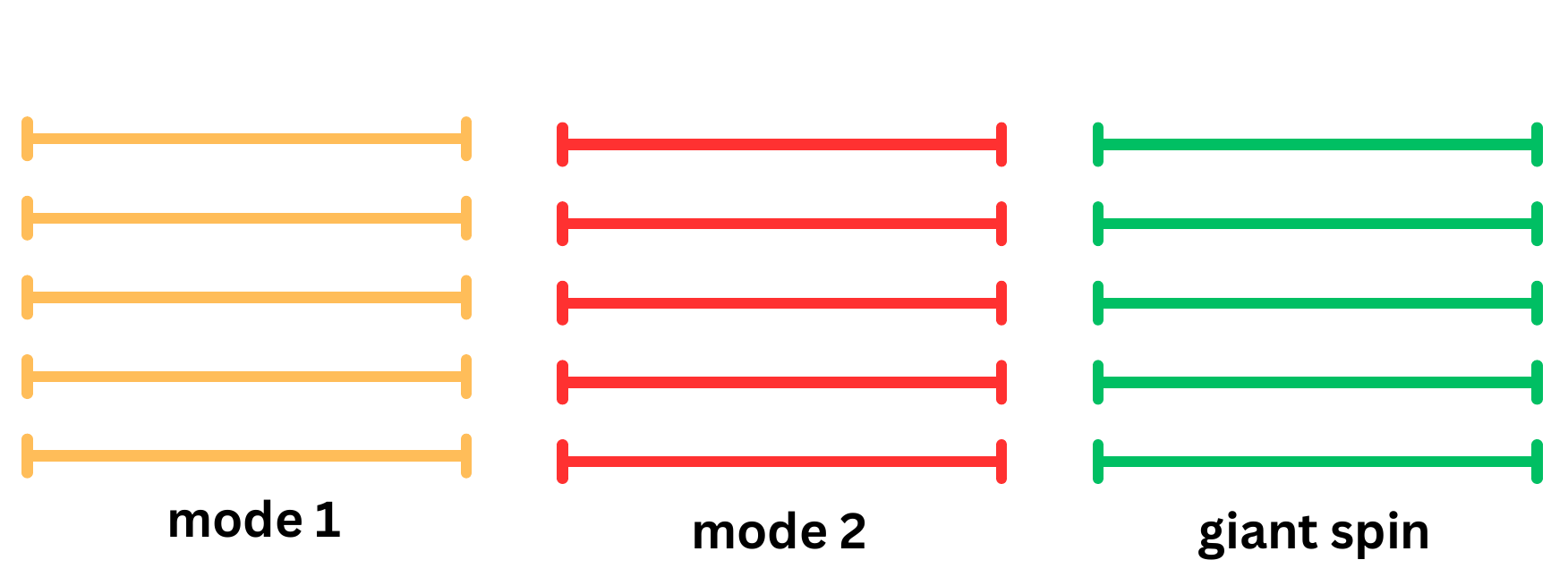}\hspace{13mm}
    \includegraphics[width=0.55\columnwidth]{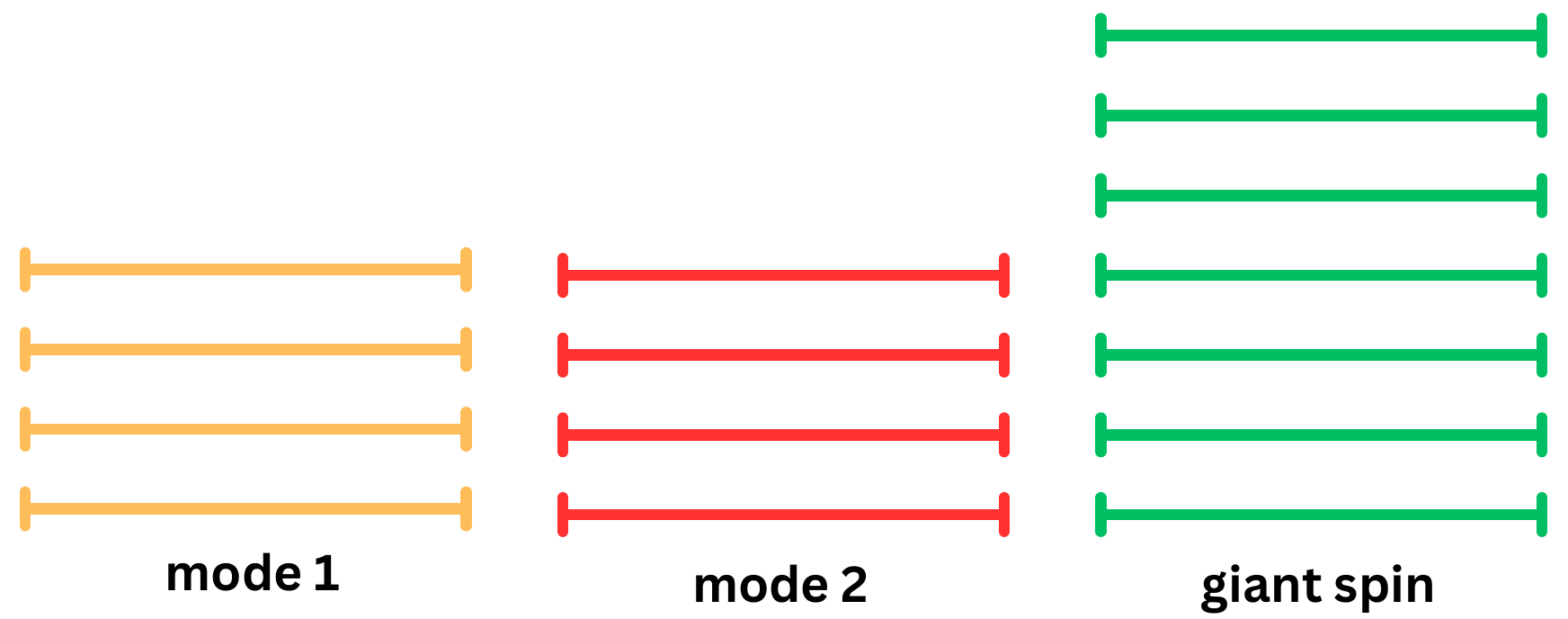}\hspace{4mm}
    \makebox[0.33\textwidth][l]{\hspace{8mm}\doubleletter[0.92]{A}}%
    \makebox[0.333\textwidth][l]{\hspace{8mm}\doubleletter[0.92]{B}}%
    \makebox[0.333\textwidth][l]{\hspace{8mm}\doubleletter[0.92]{C}}\vspace{0.4mm}\\
    \includegraphics[width=0.68\columnwidth]{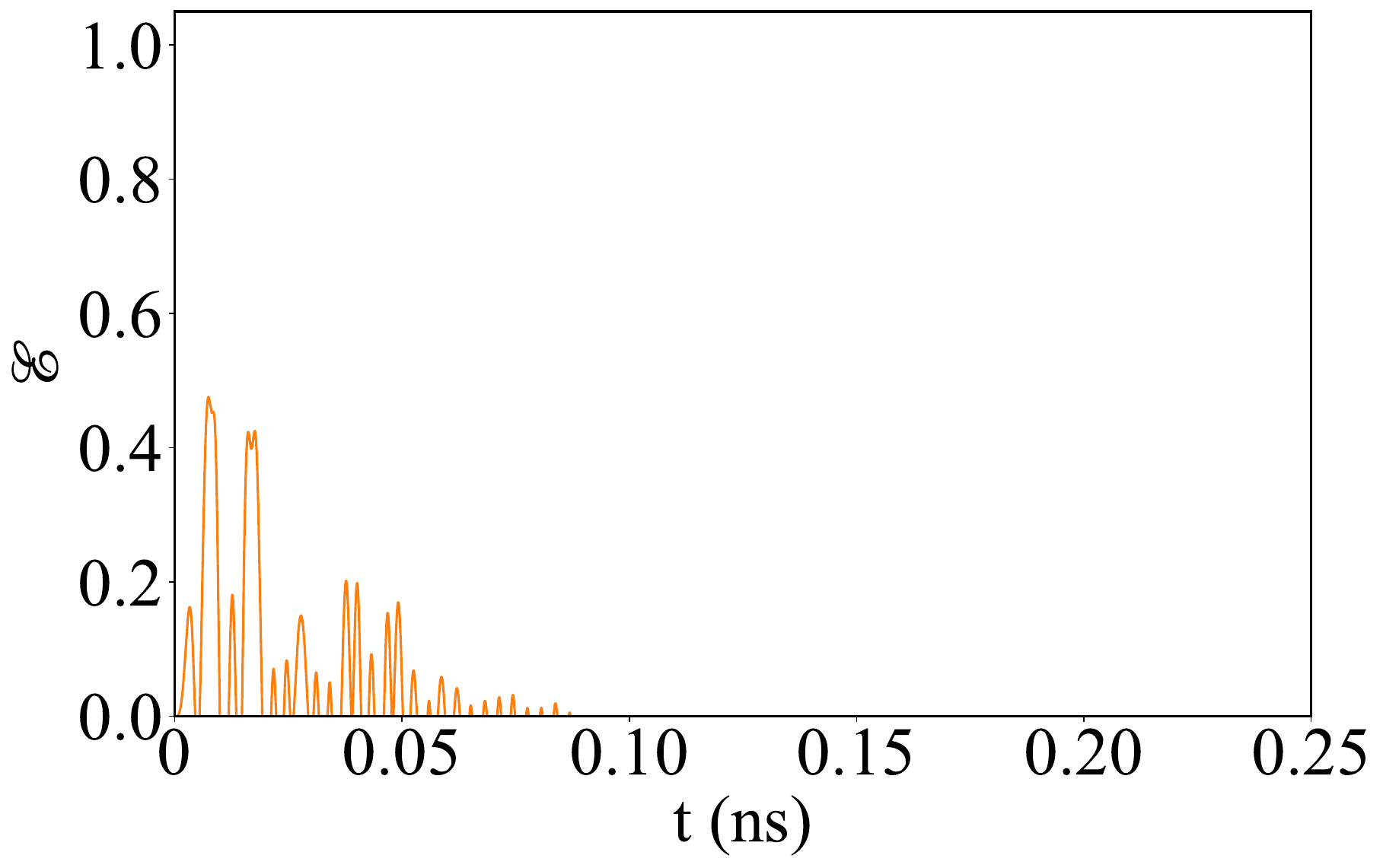}
    \hfill
    \includegraphics[width=0.68\columnwidth]{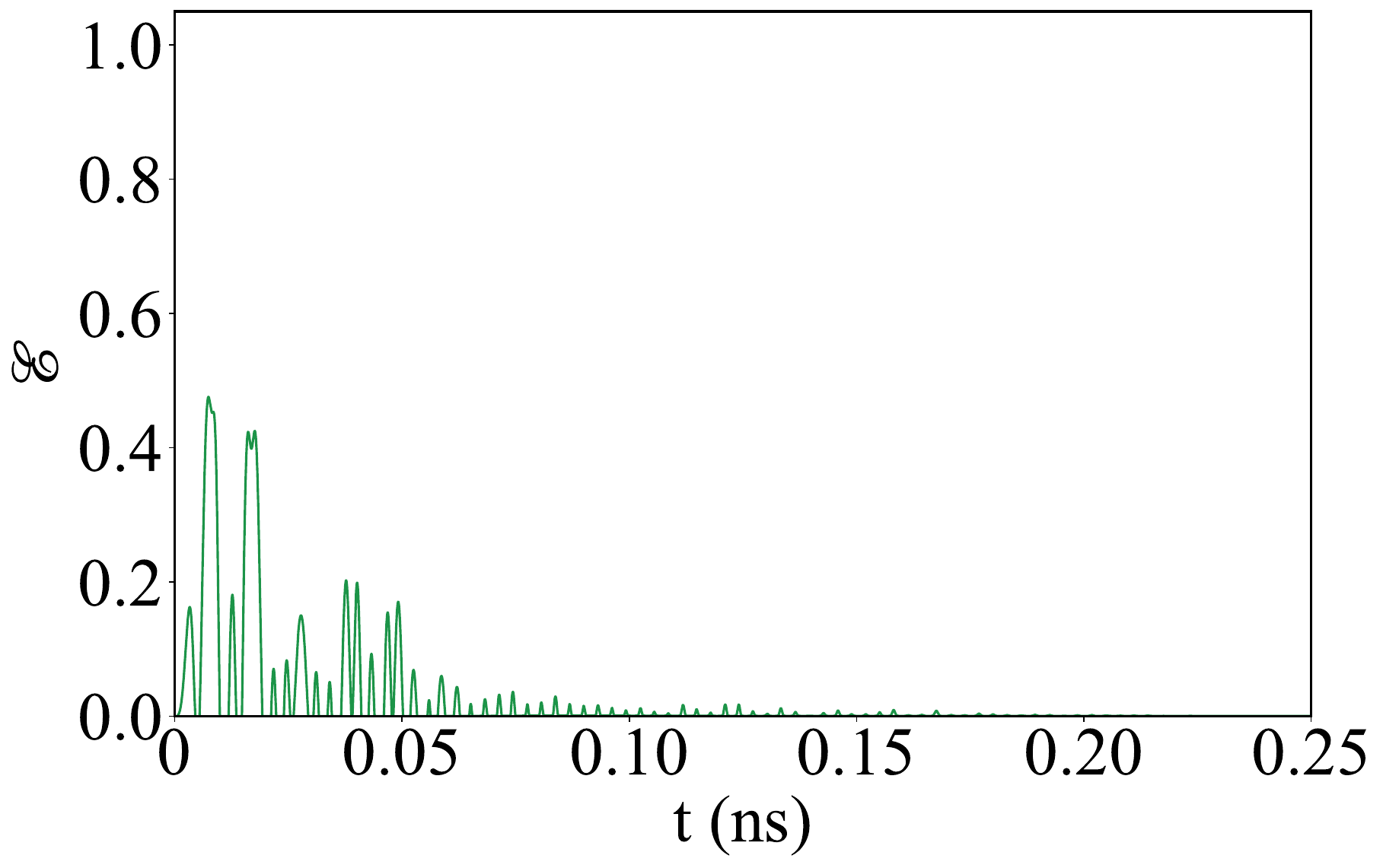}
    \hfill
    \includegraphics[width=0.68\columnwidth]
    {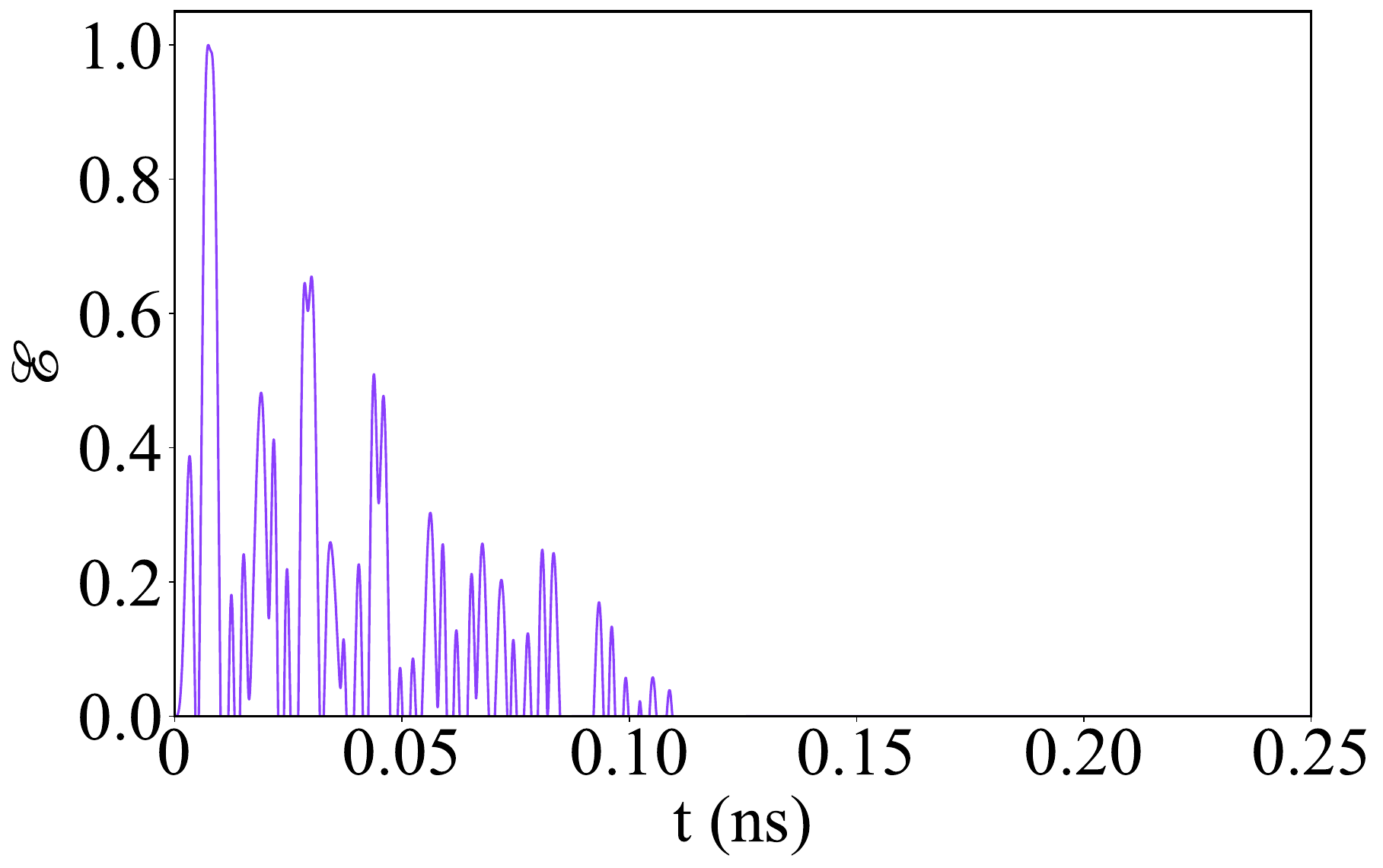}
    \captionsetup{font=small, labelfont={bf,small}}
  \caption{ Bipartite entanglement between two cavity modes interacting via a spin. \doubleletter{A.} $S=2$ and the cavity modes are four-level systems; \doubleletter{B.} $S=2$ and the cavity modes are five-level systems; \doubleletter{C.} $S=3$ and the cavity modes are four-level systems. The entanglement in each graph is renormalized with respect to the maximum entanglement in\!\doubleletter{C.} The parameters used for the simulations are: $B = 0.01~\mathrm{T}\text{,} \ E = 6.02\cdot 10^9~\mathrm{Hz}\text{,}\ D = 3.6 \cdot 10^{10}~\mathrm{Hz}\text{,} \  \omega = 6.75\cdot10^{11} ~\mathrm{Hz} \text{,} \ P = 0.01 \ \mathrm{pW} \text{,} \ \kappa_s = 10^9~\mathrm{Hz} \ \text{and} \ \kappa=7.5\cdot10^9~\mathrm{Hz}. $ }
  \label{Fig:7}
\end{figure*}
\section{Effects of dimensionality on the dynamics}
\label{appendix7}
As anticipated, our ability to quantify entanglement via the density matrix is strongly limited by computational resources. To address this issue, we start by considering a truncated Hilbert space. The truncation procedure can affect the physical consistency of the overall system, so we must ensure that our description remains physical. In the specific case, the amount of entanglement detected has to be independent of the dimension of the space chosen for the simulation.

To gain a deeper understanding, we begin by analyzing the impact of driving power. On the one hand, due to the nature of the coupling between the SMM and the photons, the driving power should not significantly alter the degree of entanglement. On the other hand, it plays an important role in the dynamical evolution of the system, specifically in determining how strongly the evolution is constrained by the boundedness of space. In Fig.~\ref{Fig:10} various simulations are shown, where all the parameters are fixed except for the driving power. According to the discussion above, we reduced the dimension from 6 total modes to 2 total modes and from infinite-level systems to finite-level systems.
The two cavity modes are now represented by four-level systems, while the giant-spin has been reduced to a five-level system ($S=2$). The full density matrix is a $80\times80$ matrix.

\noindent The expectation is that entanglement does not depend, at least not significantly and under the condition of a weakly driven cavity, on the driving power, similarly to what is observed in the covariance matrix case. Contrary to this expectation, the graphs show a detrimental effect on entanglement as the driving power increases. Namely, the more energy the lasers inject into the cavity, the lower the amount of entanglement detected. However, in the first three graphs -- \doubleletter{A}, \doubleletter{B} and \doubleletter{C} in Fig.~\ref{Fig:10} --  the amount of entanglement and its temporal profile remain consistent at short times. If the power were directly affecting the entanglement, one would expect variations even at these early times. This observation suggests that the power does not affect the \textit{amount} of entanglement mediated by the SMM, but the \textit{dynamics} of the system in a truncated space.

Indeed, the driving power is related to the number of levels of the cavity modes that become populated, and thus to the required dimension of the Hilbert space. The higher the driving, the more excited the system becomes; i.e. it requires a larger space to evolve correctly in time. If these additional levels are not included, the system cannot evolve freely and its dynamics become truncated. Based on this, we conclude that the driving power does not affect the entanglement itself, but rather the ability to properly simulate the system’s evolution. This, in turn, reflects on the amount of entanglement detected.

To test the validity of our hypothesis, we fix the power and all other parameters and observe what happens simply enlarging the space. In Fig.~\ref{Fig:7}, the two graphs on the right differ from the left ones only in terms of dimensionality.

Specifically, the graphs on the left correspond to four-level systems for the modes, while those on the right correspond to five-level systems; the levels for the giant-spin remain fixed. To facilitate visualization, a pictorial diagram of the dimension is placed above each column. 
It is interesting to point out that, for all four graphs, the entanglement at short times is the same. The differences emerge as the simulation time increases: the higher the driving power and the lower the dimensionality, the sooner the simulation yields a zero value for the entanglement.

Therefore, when moving to a higher-dimensional setting, we observe entanglement in regions where it was absent in the lower-dimensional case. This entanglement cannot be generated by simply enlarging the space, it should already have existed in the lower-dimensional simulation.  This result therefore seems to confirm what we previously stated and, in particular, it indicates that the zero entanglement observed at long times is not reliable. Therefore, from now on, we focus on the values of entanglement at short times, where the system is not yet affected by the \textit{finiteness} of the space.

Fig.~\ref{Fig:7}\!\textcolor{blue}{\doubleletter{C}} shows how the entanglement varies as the giant-spin is modified. As expected, a different giant-spin modifies the amount of the entanglement. Furthermore, while adding levels to the cavity modes does not alter the shape or the amount of the entanglement, modifying the giant-spin does impact its profile. This further confirms that the primary source of the entanglement is the nanomagnet, i.e. the giant-spin.

\clearpage
\end{document}